	\documentclass[11pt,draftcls,onecolumn,letterpaper]{IEEEtran}

	\usepackage[T1]{fontenc}			

	\usepackage{cite} 
	\usepackage{theorem}
	\usepackage{latexsym}
	\usepackage{amssymb}
	\usepackage{citesort}

	\usepackage{graphicx} 
	\graphicspath{{./Figures/}}	

	\input ./Macro/alphabet.tex	
	\input ./Macro/abrege.tex		
\def\pth#1{\left(#1\right)}                
\def\acc#1{\left\{#1\right\}}              
\def\cro#1{\left[#1\right]}

\def\bigcro#1{\bigl[#1\bigr]}

              \def\Esp#1{{\mathrm{E}}\bigcro{#1}}

\def\Exp#1{\exp\cro{#1}}
\def\Log#1{\log\cro{#1}}

\def\Erf#1{\mathrm{erf}\cro{#1}} 
\def\Erfc#1{\mathrm{erfc}\cro{#1}}
\def\Erfcx#1{\mathrm{erfcx}\cro{#1}}

\def\Pr{\mathop{\textrm{Pr}}}

\def\IF{\text{if\:}}             
             
\def\AND{\text{and\:}}

\def\FOR{\text{for\:}}

\def\WITH{\text{with\:}}

\newsavebox{\fminibox}
\newlength{\fminilength}
\newenvironment{fminipage}[1][\linewidth]
  {\setlength{\fminilength}{#1}
   \begin{lrbox}{\fminibox}\begin{minipage}{\fminilength}}
  {\end{minipage}\end{lrbox}\noindent\fbox{\usebox{\fminibox}}}

  \def\+{^\dagger}

\def\nequiv{\not\kern-.05em\equiv}
\def\egal{\kern-.5em=\kern-.5em}        
\def\propt{\kern-.2em\propto\kern-.2em} 
\def\wh#1{\widehat{#1}}                 

\def\froc#1#2{{#1/#2}}                  

\def\intdouble{\int\kern-0.3em\int}
\def\inttriple{\int\kern-0.3em\int\kern-0.3em\int}

\def\rond#1{\overset{\kern-0.33em~_\circ}{#1}}
\def\rondit[#1]#2{\overset{\kern#1~_\circ}{#2}}

\def\babs{\begin{abstract}}             \def\eabs{\end{abstract}}
\def\barr{\begin{array}}                \def\earr{\end{array}}
\def\bcc{\begin{center}}                \def\ecc{\end{center}}
\def\cl#1{\centerline{#1}}
\def\bdes{\begin{description}}          \def\edes{\end{description}}
\def\bdoc{\begin{document}}             \def\edoc{\end{document}}
\def\ben{\begin{enumerate}}             \def\een{\end{enumerate}}
\def\beqn{\begin{eqnarray}}             \def\eeqn{\end{eqnarray}}
\def\beqnl#1{\beqn\label{#1}}           \def\eeqnl#1{\label{#1}\eeqn}
\def\beqnx{\begin{eqnarray*}}           \def\eeqnx{\end{eqnarray*}}
\def\bseqn{\begin{subeqnarray}}         \def\eseqn{\end{subeqnarray}}
\def\beq#1\eeq{\begin{equation}#1\end{equation}}
\def\bal#1\eal{\begin{align}#1\end{align}}
\def\balx#1\ealx{\begin{align*}#1\end{align*}}
\def\beqx{$$}                           \def\eeqx{$$}
\def\bfig{\protect\begin{figure}}       \def\efig{\protect\end{figure}}
\def\bfigx{\protect\begin{figure*}}     \def\efigx{\protect\end{figure*}}
\def\bfigt{\protect\begin{figurette}}   \def\efigt{\protect\end{figurette}}
\def\bfl{\begin{flushleft}}             \def\efl{\end{flushleft}}
\def\bfr{\begin{flushright}}            \def\efr{\end{flushright}}
\def\bit{\begin{itemize}}               \def\eit{\end{itemize}}
\def\bmi{\begin{minipage}}              \def\emi{\end{minipage}}
\def\bfmi{\begin{fminipage}}            \def\efmi{\end{fminipage}}
\def\bpic{\begin{picture}}              \def\epic{\end{picture}}
\def\bqu{\begin{quote}}                 \def\equ{\end{quote}}
\def\bqun{\begin{quotation}}            \def\equn{\end{quotation}}
\def\bsl{\begin{slide}}                 \def\esl{\end{slide}}
\def\btabb{\begin{tabbing}}             \def\etabb{\end{tabbing}}
\def\btabl{\begin{table}}               \def\etabl{\end{table}}
\def\btablx{\begin{table*}}             \def\etablx{\end{table*}}
\def\btab{\begin{tabular}} 
\def\btabu{\begin{tabular}}             \def\etabu{\end{tabular}}
\def\btabx{\begin{tabular*}}            \def\etabx{\end{tabular*}}
\def\bbib{\begin{thebibliography}{999}} \def\ebib{\end{thebibliography}}
\def\bver{\begin{verbatim}}             \def\ever{\end{verbatim}}
\def\bca{\begin{cases}}                  \def\eca{\end{cases}}
\def\dspsty{\displaystyle}
\def\sss{\scriptscriptstyle}

	\def\ifips{\textsc{ifips}\XS}
	\def\IFIPS{Institut de Formation des Ingénieurs de Paris-Sud\XS}

	\def\ist{\textsc{ist}\XS}
	\def\IST{Information, Systèmes et Technologie\XS}

	\def\lmd{\textsc{lmd}\XS}
	\def\LMD{Licence, Master, Doctorat\XS}

	\def\Mist{Master \textsc{ist}\XS}

	\def\atis{\textsc{atis}\XS}
	\def\dess{\textsc{dess}\XS}
	\def\dea{\textsc{dea}\XS}
	\def\eea{\textsc{eea}\XS}
	\def\Fiupso{\textsc{Fiupso}\XS}
	\def\FiupsoI{\textsc{Fiupso-i}\XS}
	\def\FiupsoII{\textsc{Fiupso-ii}\XS}
	\def\FiupsoIII{\textsc{Fiupso-iii}\XS}
	\def\Maitrise{Ma\^\i trise\XS}

	\def\td{\textsc{td}\XS}
	\def\TD{travaux dirigés\XS}
	\def\GTD{Travaux dirigés\XS}

	\def\ter{\textsc{ter}\XS}

	\def\tp{\textsc{tp}\XS}
	\def\TP{travaux pratiques\XS}
	\def\GTP{Travaux pratiques\XS}

	\def\gbm{\textsc{gbm}\XS}
	\def\elq{électronique\XS}
	\def\Elq{\'Electronique\XS}
	\def\ssl{\textsc{ssl}\XS}
	\def\SSL{signaux et systèmes linéaires\XS}
	\def\GSSL{Signaux et systèmes linéaires\XS}

	\def\Unixbf{U{\footnotesize NIX}\XS}
	\def\Shellbf{S{\footnotesize HELL}\XS}
	\def\Unix{\textsc{Unix}\XS}
	\def\Linux{\textsc{Linux}\XS}
	\def\Shell{\textsc{Shell}\XS}
	\def\Korn{\textsc{Korn}\XS}
	\def\mtlb{\textsl{matlab}\XS}
	\def\Mtlb{\textsl{Matlab}\XS}
	\def\pc{\textsc{pc}\XS}

	\def\msa{\medskipamount}
	\def\Prog#1{
	\texttt{ 
	\indent\indent\indent
	\begin{minipage}{20cm}
	\begin{tabbing} 
	~~~~~~~ \= ~~~~~~~ \= ~~~~~~~ \= ~~~~~~~ \= ~~~~~~~\= ~~~~~~~ \= ~~~~~~~ \= ~~~~~~~ \= ~~~~~~~\=
	\\
	#1
	\end{tabbing}
	\end{minipage}
	}	
	}	

	\def\UPSCdO{
	\begin{small}
	\begin{tabular}{c} 
	\UPS	\\ 
	\CdO	\\ 
	-----	\\ \\ \\
	\end{tabular} 
	\end{small}
	}

	\def\Filiere#1{
	\begin{small}
	\begin{tabular}{c} 
	#1						\\ 
	Année 2006-2007	\\ 
	-----					\\ \\ \\
	\end{tabular} 
	\end{small}
	}

	\def\LogoUPS{\includegraphics[height=1.25cm,bb=0 421 596 842]{LogoCdO}}

\def\HeadUPS#1{
	\noindent
	\bcc
	\begin{tabular*}{\textwidth}{l@{\extracolsep\fill}r} 
	\UPSCdO	& \Filiere{#1}	\\ 
	\end{tabular*} 

	\vspace*{-2.5cm}
	\LogoUPS
	\ecc
}

	\def\Signature{
	\vfill
	\begin{small}
	\begin{flushright}
	\JFG \\
	Le \today
	\end{flushright}
	\end{small}
	}

	\def\Fcs{$^{\rm \,F}$\XS}
	\def\ieme{$^{\rm e}$\XS}
	\def\rib{\textsc{rib}\XS}
	\def\bp{\textsc{bp}\XS}

	\def\TSVP{
	\vfill
	\begin{flushright}
	\dots/\dots
	\end{flushright}
	\pagebreak
	}

	\def\Trait{\cl{\rule{2cm}{0.02cm}}}

	\def\BibTeX{{\rm B\kern-.05em{\sc i\kern-.025em b}\kern-.08em
				T\kern-.1667em\lower.7ex\hbox{E}\kern-.125emX}}

	\def\frem{\hfill$\triangle$}
	\def\freq{fréquence\XS}
	\def\irm{\textsc{irm}\XS}
	\def\KG{Kitagawa \& Gersch\XS}
	\def\perio{périodogramme\XS}
	\def\WV{Vigner--Ville\XS}
	\def\cpq{c'est pourquoi\XS}
	\def\Cpq{C'est pourquoi\XS}

	\def\dsp{\textsc{dsp}\XS}
	\def\tf{\textsc{tf}\XS}
	\def\fft{\textsc{fft}\XS}
	\def\ifft{\textsc{ifft}\XS}
	\def\fftdd{\textsc{fft-2d}\XS}

	\def\dimtext#1#2#3#4#5#6#7#8#9{
	\setlength{\textwidth}{#1}
	\setlength{\textheight}{#2}
	\setlength{\oddsidemargin}{#3}
	\setlength{\evensidemargin}{#4}
	\setlength{\topmargin}{#5}
	\setlength{\headheight}{#6}
	\setlength{\headsep}{#7}
	\setlength{\voffset}{#8}
	\setlength{\hoffset}{#9}
	}
\def\cro#1{\left[#1\right]}

	\def\pmu{^{-1}}

	\def\dpdx#1#2{{{\partial {#1}\over \partial {#2}}}}

	\def\est#1{\hat{#1}} 

	\def\Prob#1{\Pr{\left[#1\right]}}

	\def\fxsp#1#2{ f \left(  #1 \vert #2 \right) }

	\def\TFde#1{\Fc\cro{#1}\XS}
	\def\TLde#1{\Lc\cro{#1}\XS}
	\def\TZde#1{\Zc\cro{#1}\XS}

	\def\tf{\textsc{tf}\XS}
	\def\sf{\textsc{sf}\XS}
	\def\tl{\textsc{tl}\XS}
	\def\tz{\textsc{tz}\XS}
	\def\tpsc{\textsc{tc}\XS}
	\def\tpsd{\textsc{td}\XS}

	\def\TF{transformée de Fourier\XS}
	\def\SF{série de Fourier\XS}
	\def\TL{transformée de Laplace\XS}
	\def\TZ{transformée en $z$\XS}
	\def\TpsC{temps continu\XS}
	\def\TpsD{temps discret\XS}
	\def\RdC{région de convergence\XS}

	\def\GTF{Transformée de Fourier\XS}
	\def\GSF{Série de Fourier\XS}
	\def\GTL{Transformée de Laplace\XS}
	\def\GTZ{Transformée en $z$\XS}
	\def\GTpsC{Temps continu\XS}
	\def\GTpsD{Temps discret\XS}
	\def\GRdC{Région de convergence\XS}

	\def\VA{variable aléatoire\XS}
	\def\VAs{variables aléatoires\XS}
	\def\va{\textsc{va}\XS}

	\def\DSP{densité spectrale de puissance\XS}
	\def\dsp{\textsc{dsp}\XS}

	\def\TDS{traitement du signal\XS}
	\def\GTDS{Traitement du signal\XS}
	\def\tds{\textsc{tds}\XS}

	\def\CND{contrôle non-destructif\XS}
	\def\GCND{Contrôle non-destructif\XS}
	\def\cnd{\textsc{cnd}\XS}

\newcommand{\Entete}[5][Nom.Pr\'enom]{
\helvscript{\renewcommand{\arraystretch}{.8}
\begin{tabular}{c} Supélec
\\--------\end{tabular} \hfill
\begin{tabular}{c}Centre national de la recherche scientifique          \\
--------\end{tabular} \hfill
\begin{tabular}{c}\UPS  \\--------\end{tabular}} \\[-.4cm]%
\begin{center}\helvnormal{\textbf{LABORATOIRE DES SIGNAUX \& SYST\`EMES}}\\%
\helvfoot{Unit\'e mixte de recherche n$^{\rm o}$ 8506\\%
Sup\'elec, plateau de Moulon, 
3 rue Joliot-Curie, 91192 \sca[-3]{Gif--sur--Yvette} Cedex (France)\\%
T\'el\'ephone\,: 01 69 85 17 12 --- 
T\'el\'ecopie\,: 01 69 85 17 65 --- 
Courriel\,: {#1}@lss.supelec.fr}%
\\[0.5cm]\end{center}%
} 

\newcommand{\Inhead}[5][Name.Surname]{
\helvscript{\renewcommand{\arraystretch}{.8}
\begin{tabular}{c} Supélec
\\--------\end{tabular} \hfill
\begin{tabular}{c}Centre national de la recherche scientifique  \\
--------\end{tabular} \hfill
\begin{tabular}{c}\UPS  \\--------\end{tabular}} \\[-.4cm]%
\begin{center}\helvnormal{\textbf{LABORATOIRE DES SIGNAUX \& SYST\`EMES}}\\%
\helvfoot{Unit\'e mixte de recherche n$^{\rm o}$ 8506\\%
Sup\'elec, plateau de Moulon, 
3 rue Joliot-Curie, 91192 \sca[-3]{Gif--sur--Yvette} Cedex (France)\\%
Telephone: 01 69 85 17 12 --- 
Fax: 01 69 85 17 65 --- 
E-mail: {#1}@lss.supelec.fr}%
\\[0.5cm]\end{center}%
}

	  \def\PDB{\gamma_\dD}    \def\PD{r_\dD}    
	  \def\PAB{\gamma_\bD}    \def\PA{r_\bD}    
	  \def\PNB{\gamma_\nD}    \def\PN{r_\nD}    
	  \def\PDNAB{\gammab}	  

		\def\Datab{\Yb}	\def\Data{Y}	\def\Datac{\Yc}	\def\data{y}
		\def\Auxb{\Bb}		\def\Aux{B}		\def\Auxc{\Bc}		\def\aux{b}
		\def\Objb{\Xb}		\def\Obj{X}		\def\Objc{\Xc}		\def\obj{x}
		\def\Bruitb{\Nb}	\def\Bruit{N}	\def\Bruitc{\Nc}	\def\bruit{n}
		\def\MatHb{\Hb}	\def\MatH{H}	\def\MatHc{\Hc}	\def\matH{h}
		\def\MatDb{\Db}	\def\MatD{D}	\def\MatDc{\Dc}	\def\matD{d}
		\def\MatFb{\Fb}	\def\MatF{F}	\def\MatFc{\Fc}	\def\matF{f}
		\def\Moyb{\Mb}		\def\Moy{M}		\def\Moyc{\Mc}		\def\moy{m}
	
		\def\InvErf#1{\mathrm{ierf}\acc{#1}}
		\def\erf{\mathrm{erf}}
		\def\erfc{\mathrm{erfc}}
		\def\inverf{\mathrm{ierf}}
		\def\erfcx{\mathrm{erfcx}}

		\def\eap{\textsc{eap}\XS}
		\def\fftdd{\textsc{fft-2d}\XS}

		\def\eps{\varepsilon}
		\def\phi{\varphi}
		\def\Schur{\otimes}
		\def\PostMean{_\textrm{PM}}
		\def\CondPostMean{_\textrm{CPM}}
		\def\MAP{_\textrm{MAP}}

		\newtheorem{remark}{Remark}
		\newtheorem{property}{Property}
		\newtheorem{example}{Example}

\begin{document}

%

\title{Unsupervised bayesian convex deconvolution  based on a field with an explicit partition function}

\author{J.-F. Giovannelli\thanks{J.-F.~Giovannelli is with Laboratoire des Signaux et Syst\`emes (\textsc{cnrs} -- Sup\'elec -- \textsc{ups}), Sup\'elec, Plateau de Moulon, 3 rue Joliot-Curie, 91192 Gif-sur-Yvette Cedex, France. E-mail: giova@lss.supelec.fr.}}

\markboth{Submitted to IEEE TRANSACTIONS ON IMAGE PROCESSING}{GIOVANNELLI: UNSUPERVISED BAYESIAN CONVEX DECONVOLUTION}

\maketitle

\begin{abstract}
This paper proposes a non-Gaussian Markov field with a special feature: an \textit{explicit} partition function. To the best of our knowledge, this is an original contribution. Moreover, the explicit expression of the partition function enables the development of an \textit{unsupervised} edge-preserving convex deconvolution method. The method is fully Bayesian, and produces an estimate in the sense of the posterior mean, numerically calculated by means of a Monte-Carlo Markov Chain technique. 
%
%
%
The approach is particularly effective and the computational practicability of the method is shown on a simple simulated example. 
\end{abstract}

\medskip 

\begin{keywords}
Deconvolution, Bayesian statistics, regularization, convex potentials, partition function, hyperparameters estimation, unsupervised estimation, Monte-Carlo Markov Chain.
\end{keywords}
 
%
\section{Introduction}

The research concerning regularization for ill-posed inverse problems was first carried out by Phillips, Twomey and Tikhonov~ in the sixties and are compiled in~\cite{Tikhonov77}. For the specific problem of deconvolution they lead to the contributions of Hunt~\cite{Andrews77} based on toroidal models and fast implementation by \fft. These methods rely on quadratic penalization \ie Gaussian laws in a Bayesian framework. The solutions thus formulated are linear \wrt the data and numerically efficient. However, their resolution is limited: the capability to properly restore sharp edges is limited.

At the beginning of the eighties, in order to overcome these limitations, Geman \& Geman~\cite{Geman84} (see also~\cite{Blake87}) introduced a much superior Markovian field including hidden variables~\cite{Idier99}. The hidden variables (also referred to as dual or auxiliary variable) are binary and interactive variables modeling sharp edges and closed contours. The data processing then relies on a detection-estimation strategy and allows the recovery of distinct zones with abrupt changes. The calculation of the solution in the sense of the maximum \apost is based on a simulated annealing algorithm which requires intensive numerical computations. For the sake of computational efficiency in some cases, Geman \& Reynolds~\cite{Geman92} and then Geman \& Yang~\cite{Geman95} introduced auxiliary (also referred to as dual) variables: the sampling of a correlated non-Gaussian field reduces to the sampling of a correlated Gaussian field for one part and to the sampling of a separable field for the other. Furthermore, the construction proposed by~\cite{Geman95} is founded on the work of Hunt and the toroidal models: the sampling of the correlated Gaussian field reduces to the sampling of an inhomogeneous white Gaussian field followed by an \fft. The proposal below takes advantage of this construction.


The case of fields with convex potential~\cite{Bouman93,Green90,Rudin92,Kunsch94,OSullivan95,Fessler00} (see also~\cite{Figueiredo03,Starck01}) was laid down in the nineties as fulfilling a compromise between the quality of the reconstructed images and the computational burden. In this framework, a particular attention has been paid to the case of $\mbox{L}_2-\mbox{L}_1$ potentials~\cite{Green90,Rudin92,Kunsch94,OSullivan95,Fessler00}:  a quadratic behavior around the origin and a linear behavior at large values allow edge preservation. In this context, the constructions of~\cite{Geman92} and~\cite{Geman95} respectively led to two algorithms: ARTHUR and LEGEND~\cite{Charbonnier97} (see also~\cite{Idier01b}). The work presented here concerns this type of potential.



With such potentials, the regularized solutions usually necessitate the adjustment of three hyperparameters: two parameters to control the law for the image and one parameter to control the law for the noise. Several attempts are dedicated to the question of  hyperparameter estimation and the investigated solutions are frequently based on statistical approaches: (approximated or pseudo) likelihood, Bayesian strategies, EM and SEM algorithms\dots The reader may consult papers such as~\cite{Zhou97,Figueiredo97,Saquib98,Descombes99a,Molina99,Lanterman00,Pascazio03} and reference books such as~\cite[Part.VI]{Winkler03}, \cite[Ch.7]{Li01} or~\cite[Ch.8]{Idier01a}. These approaches are potentially very powerful but they come up against a major difficulty: the partition function of existing \aprio fields depends on hyperparameters and is not explicitly given. 

The first novelty of the paper lies in the fact that it proposes a new random field with an explicit partition function. To this end, the paper build an original type of compound (toroidal) field with $\mbox{L}_2-\mbox{L}_1$ potential. 
%
%
The work is largely inspired by the Bayesian interpretation of dual variables in terms of Location Mixture of Gaussian  proposed by~\cite{Champagnat04}. Moreover, it is also inspired by~\cite{Jalobeanu02} (itself based on the contributions of Hunt~\cite{Andrews77} and Geman \& Yang~\cite{Geman95}). However, none of these contributions put forward the idea of a field with an explicit partition function. 
%
%
%
Afterwards, the paper proposes a second novelty: a full Bayesian unsupervised (\ie including hyperparameter estimation) edge-preserving convex deconvolution method, thanks to the knowledge of the partition function.
%
%
It is based on a \post law for the whole set of unknown parameters (including hyperparameters) and a Minimum Mean Square Error strategy. 

The paper is presented in the following manner. Section~\ref{Sec:Statement} introduces the notations and states the problem.  Section~\ref{Sec:Prior} is devoted to the construction of the proposed field, Section~\ref{Sec:Deconvolution} proposes its use for image deconvolution and demonstrates the numerical practicability. Conclusions and perspectives are delivered in Section~\ref{Sec:Conclu}. Most of the calculations are explained in Appendices~\ref{Sec:AnnErf} to~\ref{Sec:AnnInitHyper}.


\section{Notation and problem statement} \label{Sec:Statement}


Work is carried out on $P\times P$ real images, with $N=P^2$ pixels, represented in a matrix form. $a_{pq}$ denotes the generic element of the matrix $\Ab$, $N_2(\Ab) = \sum |\,a_{pq}\,|^2$ its squared norm and $\rond{\Ab}$ its \fftdd. 
The transformation is normalized: the Parseval relationship is written as $N_2(\Ab)=N_2(\rond{\Ab})$ and the sum of the pixels is $\sum a_{pq} = P\,\rond{a}_{\scriptscriptstyle 00}$.  
The symbols  $\star$ and $\Schur$ respectively represent the circular convolution and the Schur product (termwise) of matrices. If $\MatFb$ represents a circular filter and $\Ib$ an input object, the output is written $\Ob=\MatFb\star\Ib$ in the spatial domain resulting in $\rond{\Ob}=\rond{\MatFb}\Schur\rond{\Ib}$ in the Fourier domain. If $\rond{\matF}_{pq}\neq 0$ for all $p,q$, the associated filter is invertible. 

In the subsequent developments about deconvolution, $\Datab$, $\Objb$, $\MatHb$, and $\Bruitb$ respectively denote the observed data, the unknown object, the convolution matrix and the observation noise. With these notations, the observation equation is written:
\beq \label{Eq:Convol}
\Datab = \MatHb \star \Objb + \Bruitb \,.
\eeq
The deconvolution problem consists in recovering the unknown object $\Objb$ given the observed data $\Datab$ and given the observation model $\MatHb$. 
The ill-posedness of the problem has been well identified for several decades and the problem is nowadays often tackled in a Bayesian framework using Markov priors. In a Gibbs form, the prior law writes:
\beqx
f_{\Objc}\cro{\Objb}  = K_{\Objc}\pmu ~ \Exp{ - \Phi_\thetab \pth{\Objb} }  \,,
\eeqx
where $K_{\Objc}$ is the partition function (normalizing constant) and $\Phi_\thetab$ is the Gibbs energy controlled by a set of parameters (such as variance, threshold, scale, correlation length\dots) collected in a vector $\thetab$. The general methodology is well-known: the solution is determined from the \apost law and a point estimate can be chosen as the mean or the maximizer, for instance. 
%
%
%
Anyway, the posterior law (and the point estimates) depends upon hyperparameters notably on the parameters of the prior~$\thetab$.  The inference about these parameters can be attempted in a statistical framework whose keystone is an exact and explicit likelihood function (in an usual sense or in a posterior sense). This function is itself founded on a complete expression for the prior law including the partition function as it depends on~$\thetab$. It is given as a large dimension integral:
\beq  \label{Eq:Partition}
K_{\Objc} = K_{\Objc}(\thetab) = \displaystyle \int_{\eR^N} \Exp{ \, - \Phi_\thetab \pth{\Objb} \, } ~\dD \Objb \,.
\eeq
%






It is a commonplace to say that $K_{\Objc}(\thetab)$ can be explicitly given for two well-known classes of (continuous state) field: 
\bit

\item[($i$)] $\Phi$ is quadratic, \ie the field is Gaussian

\item[($ii$)] $\Phi$ is separable, \ie the field is white. 

\eit
%
%
%
%
%
%
In other cases and especially for non-separable and non-Gaussian fields, the theoretical calculation and the numerical computation of (\ref{Eq:Partition}) are desperate tasks~\cite[p.281]{Winkler03} and they have never been achieved\footnote{The partition function is however known for the Ising field~\cite{Onsager44}. It is a binary field out of the scope of the developed work.}. 
However, its achievement is made possible and simple in the next Section, for a special non-separable  and non-Gaussian field. 
%

%
%
%
%
%

\section{Prior Field with Partition Function} \label{Sec:Prior}

Taking advantage of ($i$) and ($ii$) above, the proposed random field is a compound field involving two variables: a pixel variable noted as $\Objc$ and an auxiliary (or dual or hidden) variable noted as $\Auxc$. The joint law for $(\Objc,\Auxc)$ is defined by the law of $\Objc|\Auxc$ for one part and by the law of $\Auxc$ for the other part. The former is a Gaussian component (case ($i$) above) and the latter is a separable component (case ($ii$) above). 

%
%



%



\subsection{Toroidal Gaussian Field for $\Objc|\Auxc$}

Let us consider two matrices $\Auxb$ and $\MatFb$ with $\rond{\matF}_{pq}\neq 0$ for all $p,q$ and the toroidal (circular shift invariant) Gaussian field with a density parametrized in the form:
\beq \label{Eq:ObjCondAux}
f_{\Objc|\Auxc}\cro{\Objb|\Auxb}  = K_{\Objc|\Auxc}\pmu ~ \Exp{ - \PDB \,  N_2( \MatFb \star \Objb - \Auxb) /2 }  \,,
\eeq
where $\PDB>0$ is an inverse variance. The matrix $\MatFb$ designs the field structure and especially the neighborhood system and the form of the cliques. In the Fourier domain, the potential is separable and naturally develops in two forms: 
\beqnx
N_2(\MatFb\star\Objb-\Auxb) &=& N_2\pth{ \rond{\MatFb}  \Schur \rond{\Objb}-\rond{\Auxb}  } \\
										&=& \sum_{pq} | \, \rond{\matF}_{pq}  \, \rond{\obj}_{pq}-\rond{\aux} _{pq} \, |^2 \\
										&=& \sum_{pq} | \, \rond{\matF}_{pq} |^2 ~ | \, \rond{\obj}_{pq} - \rond{\aux} _{pq} / \rond{\matF}_{pq} \, |^2
\eeqnx
which has three essential consequences for the following developments.

\ben

\item The law for $\rond{\Objc}$ is separable and each $\rond{\Obj}_{pq}$ is Gaussian with mean $\rond{\aux}_{pq} / \rond{\matF}_{pq}$  and  inverse variance $\PDB |\rond{\matF}_{pq}|^2$. As a result, the sampling of $\Objc$ reduces to the sampling of an inhomogeneous white Gaussian noise followed by an \fftdd.

\item The change of variable $\overline{\Objc} = \MatFb \star \Objc$ is invertible, $\overline{\Objc}$ is white and each $\overline{\Obj}_{pq}$ is Gaussian with mean $\aux_{pq}$  and common inverse variance  $\PDB$.

\item The partition function $K_{\Objc|\Auxc}$ is easily tractable in the Fourier domain thanks to a change of variable 
\beqnx \label{Eq:PartGauss}
K_{\Objc|\Auxc} &=&  \displaystyle \int_{\eR^N} \Exp{ - \PDB \,  N_2( \MatFb \star \Objb - \Auxb) /2  }  ~\dD \Objb \\
                    &=& \cro{ \PDB^{\,N/2} ~~  \pth{2\pi}^{-N/2} ~ \prod |\rond{\matF}_{pq}| }\pmu \,,
\eeqnx 
and does not depend on $\Auxb$.

\een

In relation to existing works such as \cite{Geman95,Charbonnier97,Idier01a,Jalobeanu02,Champagnat04}, the main idea here is simply to focus on the case where the change of variable $\overline{\Objc} = \MatFb \star \Objc$ is invertible (point 2 above) that is to say the number of cliques and the number of pixels are equal.


\begin{remark}---~
%
%
The partition function $K_{\Objc|\Auxc}$ does not depend on $\Auxb$ as a counterpart of a limitation: the number of cliques and the number of pixels are equal. As an illustration of the limitation, let us point out that $K_{\Objc|\Auxc}$ depends on $\Auxb$ for a field based on horizontal cliques plus vertical cliques (the number of cliques is greater than the number of pixels). 
\end{remark}


\begin{figure}[htb]
\bcc\btabu{c}
\includegraphics[width=4.5cm]{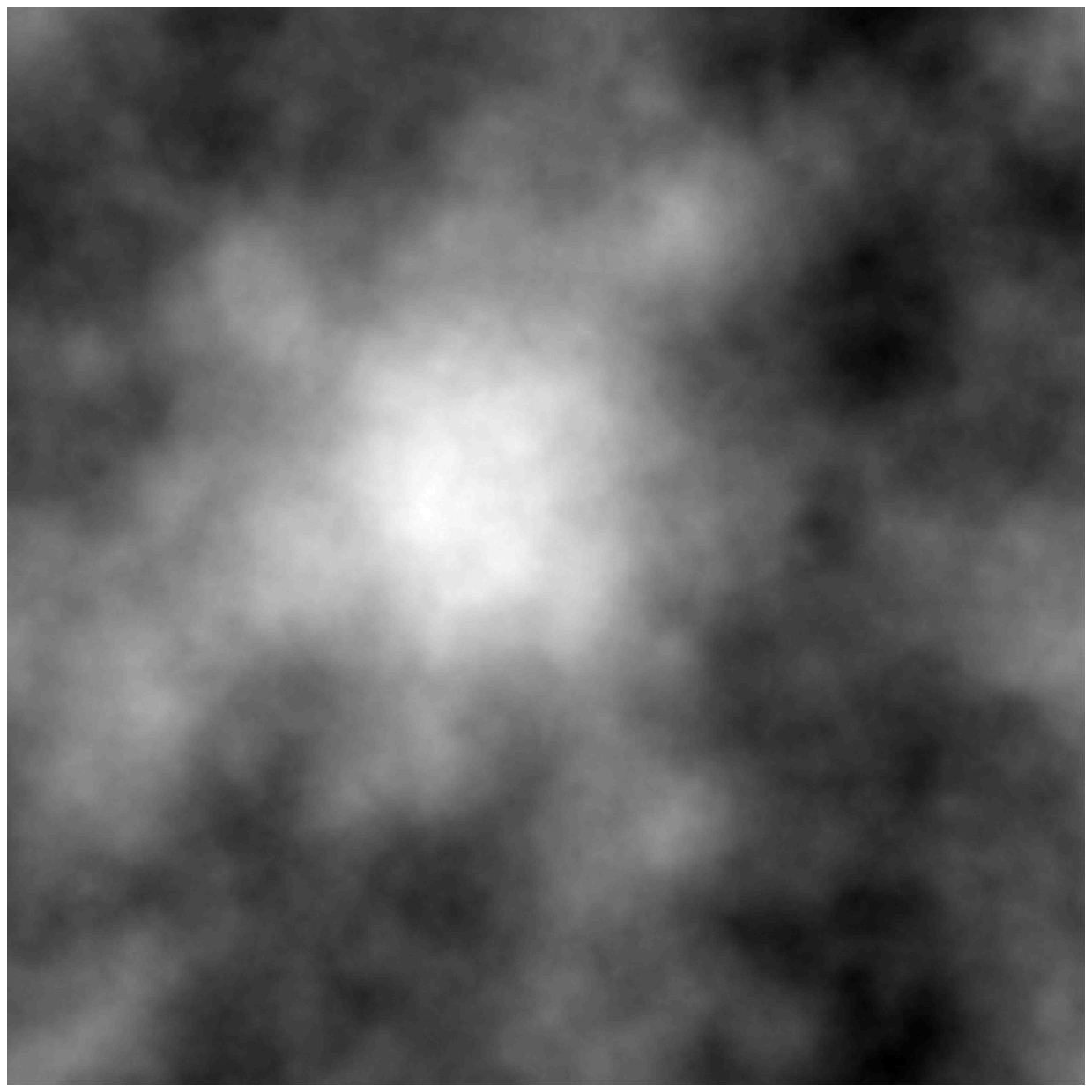}\\
\etabu\ecc
\caption{A sample of the field, with $\PDB=\PAB=1$ ($\eps$ is also set to 1).\label{Fig:UnPrior}}
\end{figure}

\begin{figure}[htb]
\bcc\btabu{c}
~~\includegraphics[width=7cm]{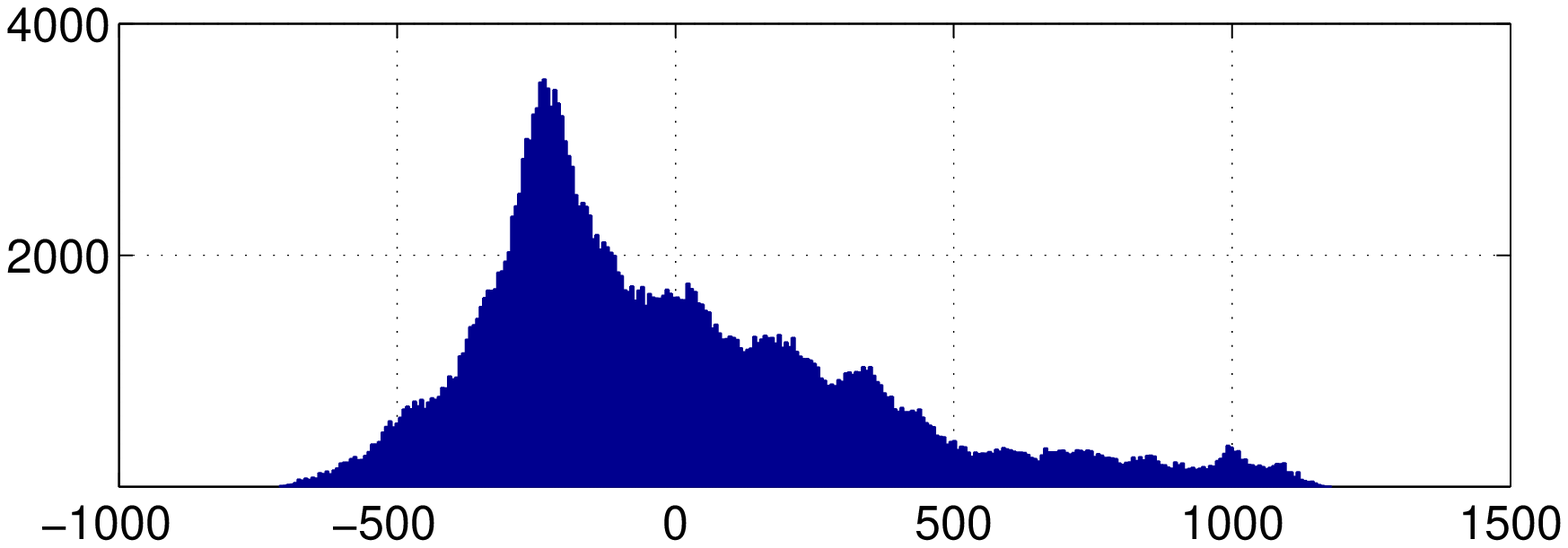}\\
\includegraphics[width=7cm]{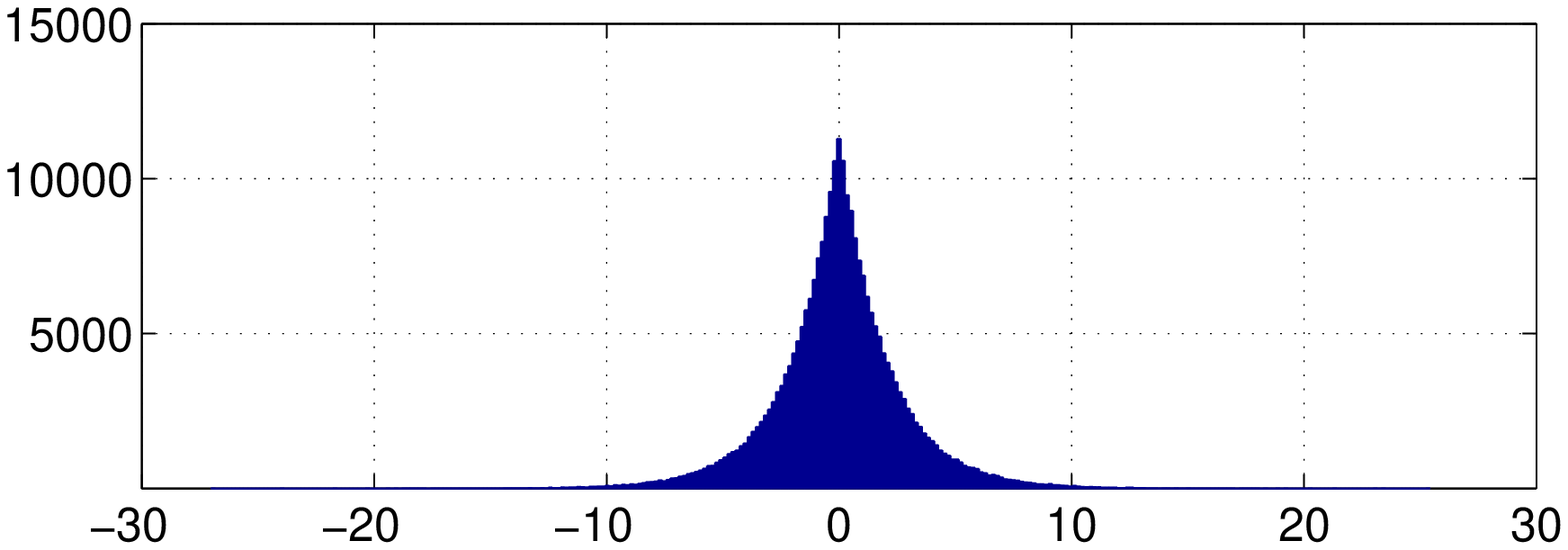}\\
\includegraphics[width=7cm]{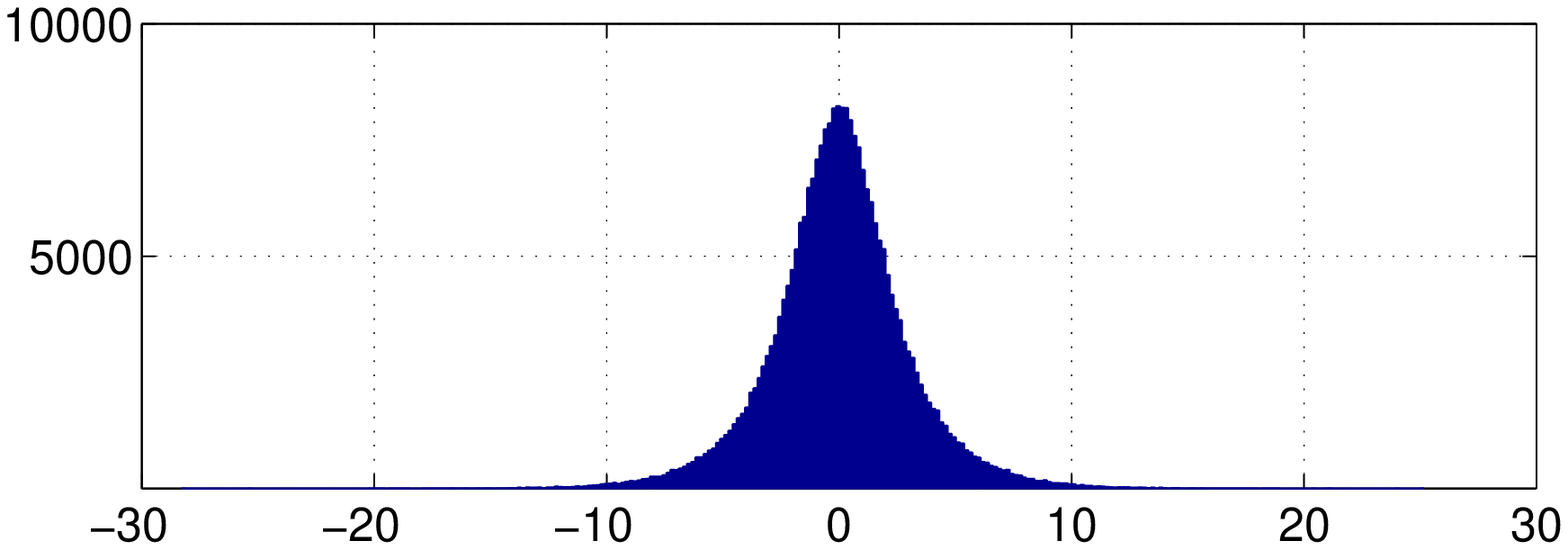}\\
\etabu\ecc
\caption{Histograms (image at Fig.~\ref{Fig:UnPrior}). From top to bottom: histogram of image pixels $\Obj$, histogram of auxiliary variables $\Aux$ and histogram of differences $\overline{\Obj}$.\label{Fig:Histo}}
\end{figure}

\subsection{Compound Field} \label{Sec:Compound}

A separable and homogeneous field is then introduced for the auxiliary variable $\Auxc$ with a density $f_{\Auxc}\cro{\Auxb}$, product of the  $f_{\Aux}\cro{\,\aux_{pq}\,}$. The joint density is written as $f_{\Objc,\Auxc}\cro{\Objb,\Auxb} =  f_{\Objc|\Auxc}\cro{\Objb|\Auxb} f_{\Auxc}\cro{\Auxb}$ and the marginal law is obtained by integrating the auxiliary variables:
\beqx 
 f_{\Objc}\cro{\Objb} = \displaystyle \int_{\eR^N} f_{\Objc|\Auxc}\cro{\Objb|\Auxb} f_{\Auxc}\cro{\Auxb} ~\dD \Auxb \,.
\eeqx
Since the partition function $K_{\Objc|\Auxc}$ does not depend on $\Auxb$, the calculations can be achieved
\beqx 
\barr{l} 
~\\
~~ f_{\Objc}\cro{\Objb}    
~\\
~\\
=\displaystyle K_{\Objc|\Auxc}\pmu ~ \int_{\eR^N} f_{\Auxc}\cro{\Auxb} \, \Exp{ - \, \PDB \,  N_2( \MatFb \star \Objb - \Auxb) \,/2}   \,  ~\dD \Auxb
~\\
~\\
=\displaystyle K_{\Objc|\Auxc}\pmu ~ \prod_{pq} \int_{\eR} f_{\Aux}\cro{\,\aux_{pq}\,} \, \Exp{ - \, \PDB \, \pth{\overline{\obj}_{pq}-\aux_{pq}  }^2 \,/2  } ~\dD \aux_{pq}
\earr
\eeqx
which involves a separable convolution product. 

\begin{remark}---~
The proposed construction is possible for any probability density function $f_{\Aux}$. In this sense, it is possible to design a large class of potential functions. 
\end{remark}

Thus, a wide range of law is available, but the convex potential case is the one of interest here, as mentioned in the introduction. So, the following property is of importance.

\begin{property}---~\label{Rem:CVX}
For any log-concave probability density function $f_{\Aux}$, the probability density function $f_{\Objc}$ is log-concave~\cite[Theo.~7]{Prekopa73},\cite{Ibragimov56}.
\end{property}





\begin{figure}[htb]
\bcc\btabu{r}
\includegraphics[width=8.25cm]{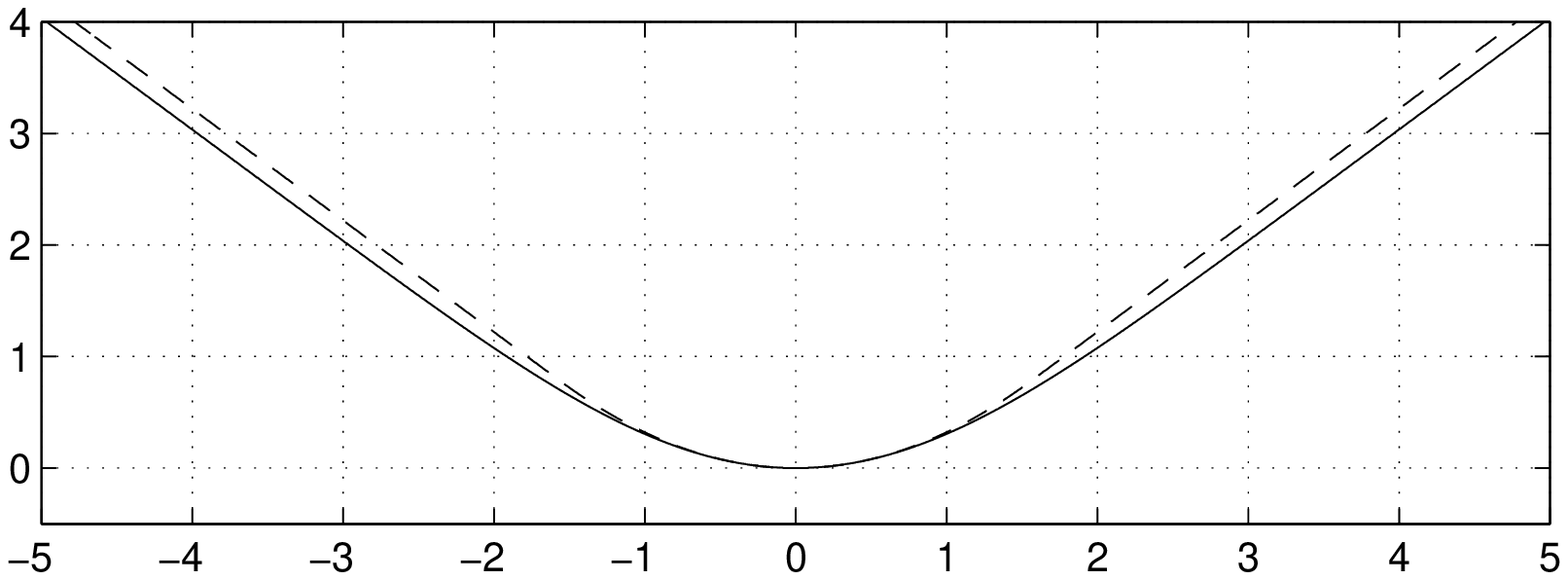}\\
\includegraphics[width=8.5cm]{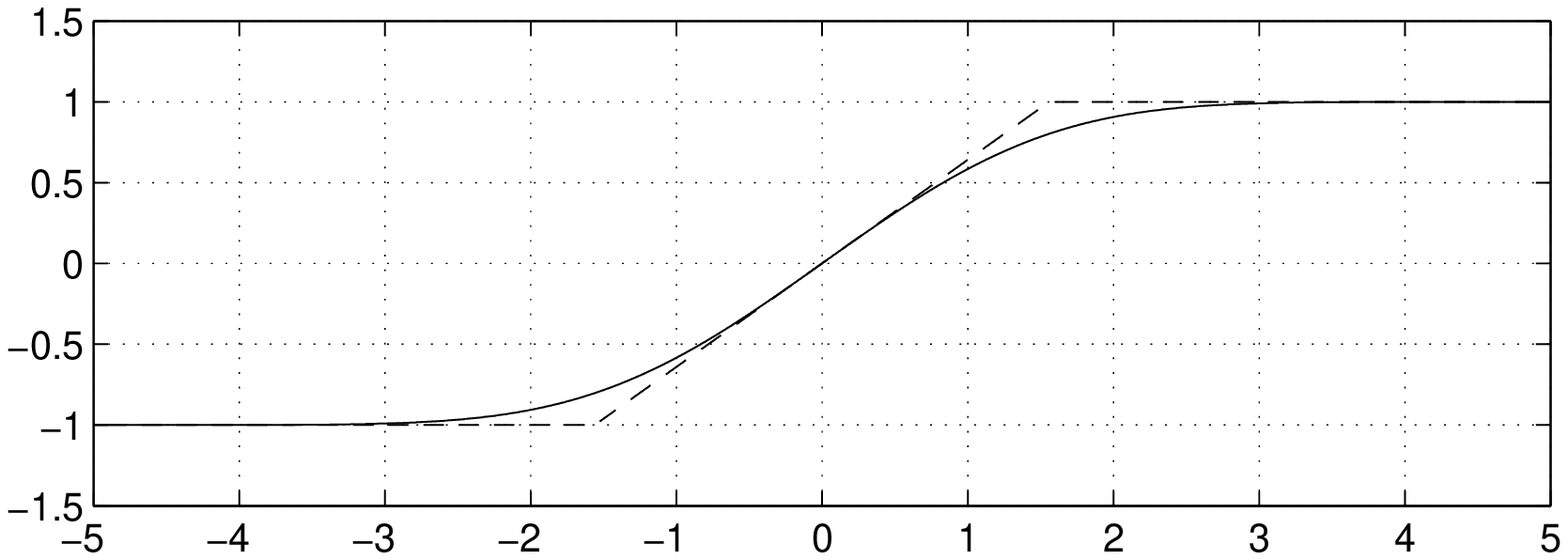}\\
\includegraphics[width=8.5cm]{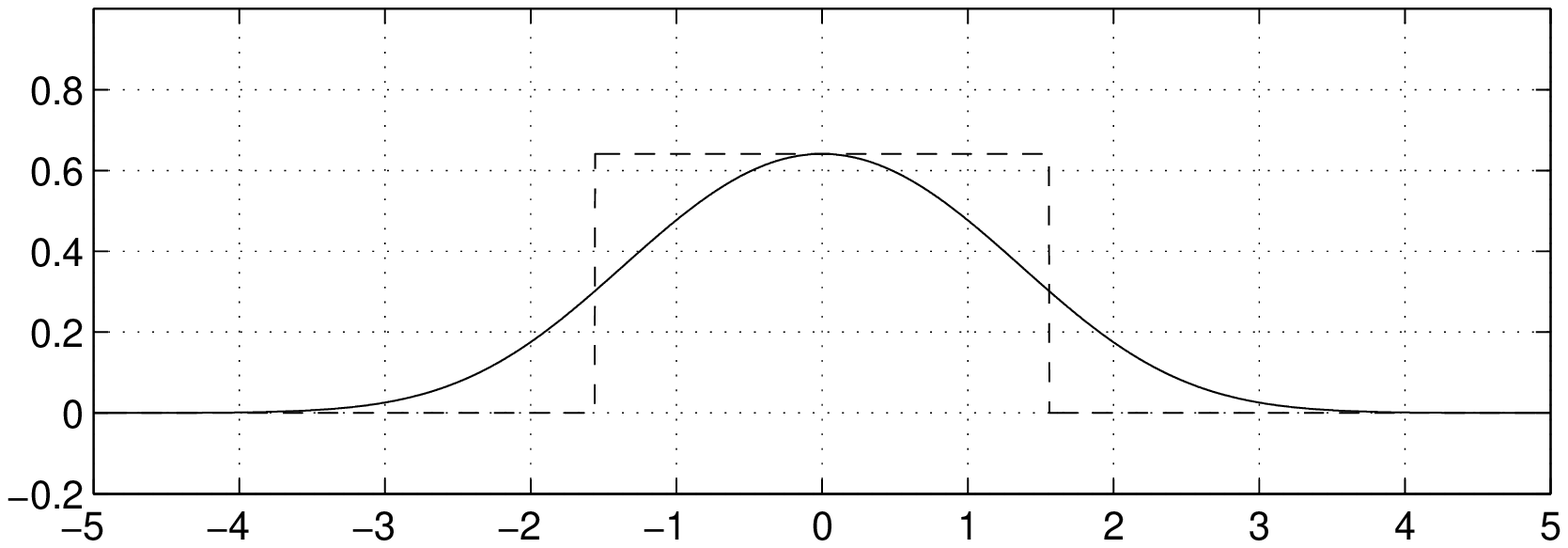}\\
\etabu\ecc
\caption{From top to bottom: potential function, first and second derivative. Solid line: Log-Erf potential $\phi(x)$ of Eq.~(\ref{Eq:PotentielLogErf}) and dotted line: corresponding Huber potential of Eq.~(\ref{Eq:PotentielHuber}). The potential parameters are  $\PDB=\PAB=1$ and hence the equivalent Huber parameters are $\lambda\simeq0.32$ and $s\simeq1.56$, according to Eq.~(\ref{Eq:LambdaSeuil}). \label{Fig:Potentiel}}
\end{figure}

\begin{figure}[htb]
\bcc
\includegraphics[width=4.25cm]{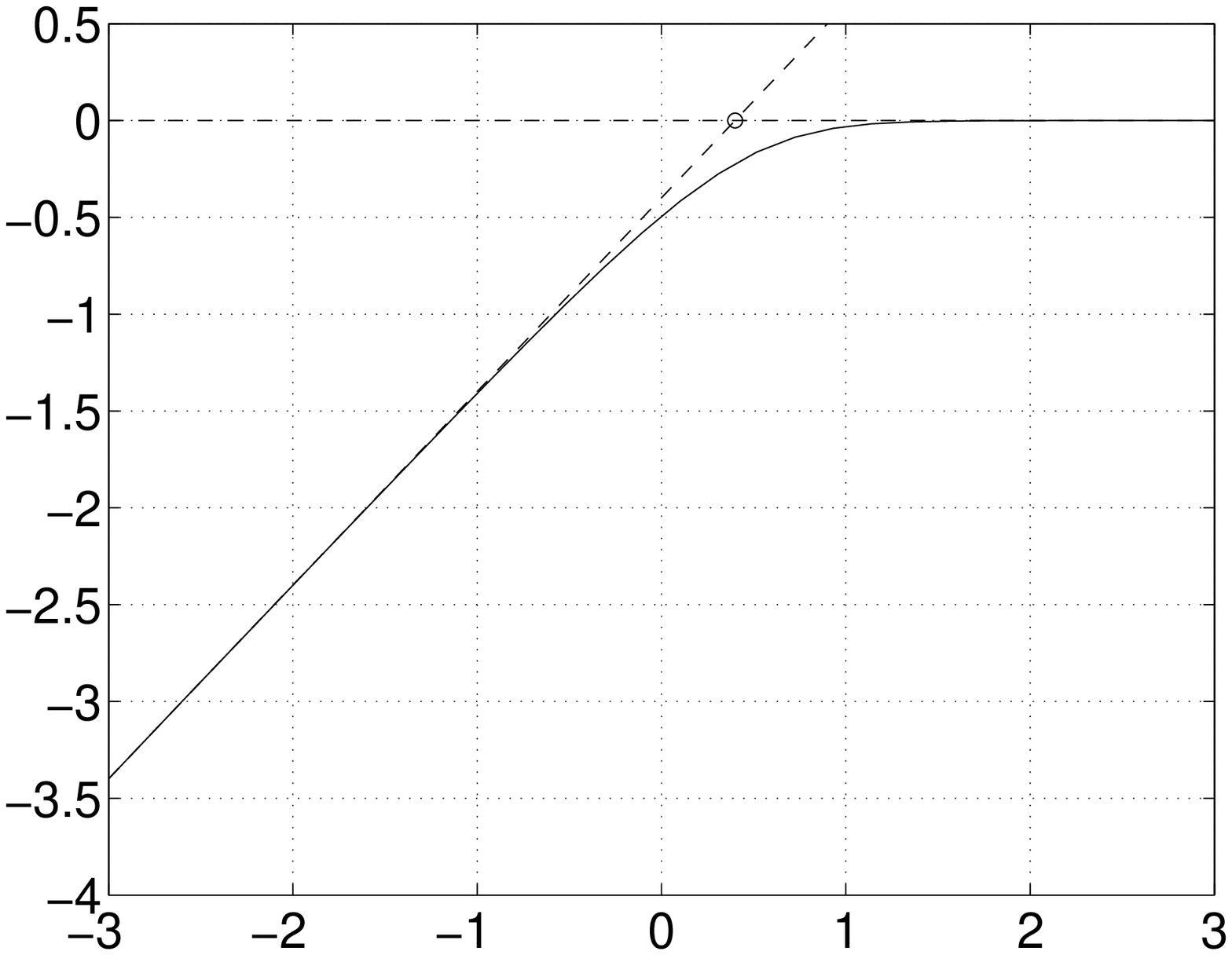} \includegraphics[width=4.25cm]{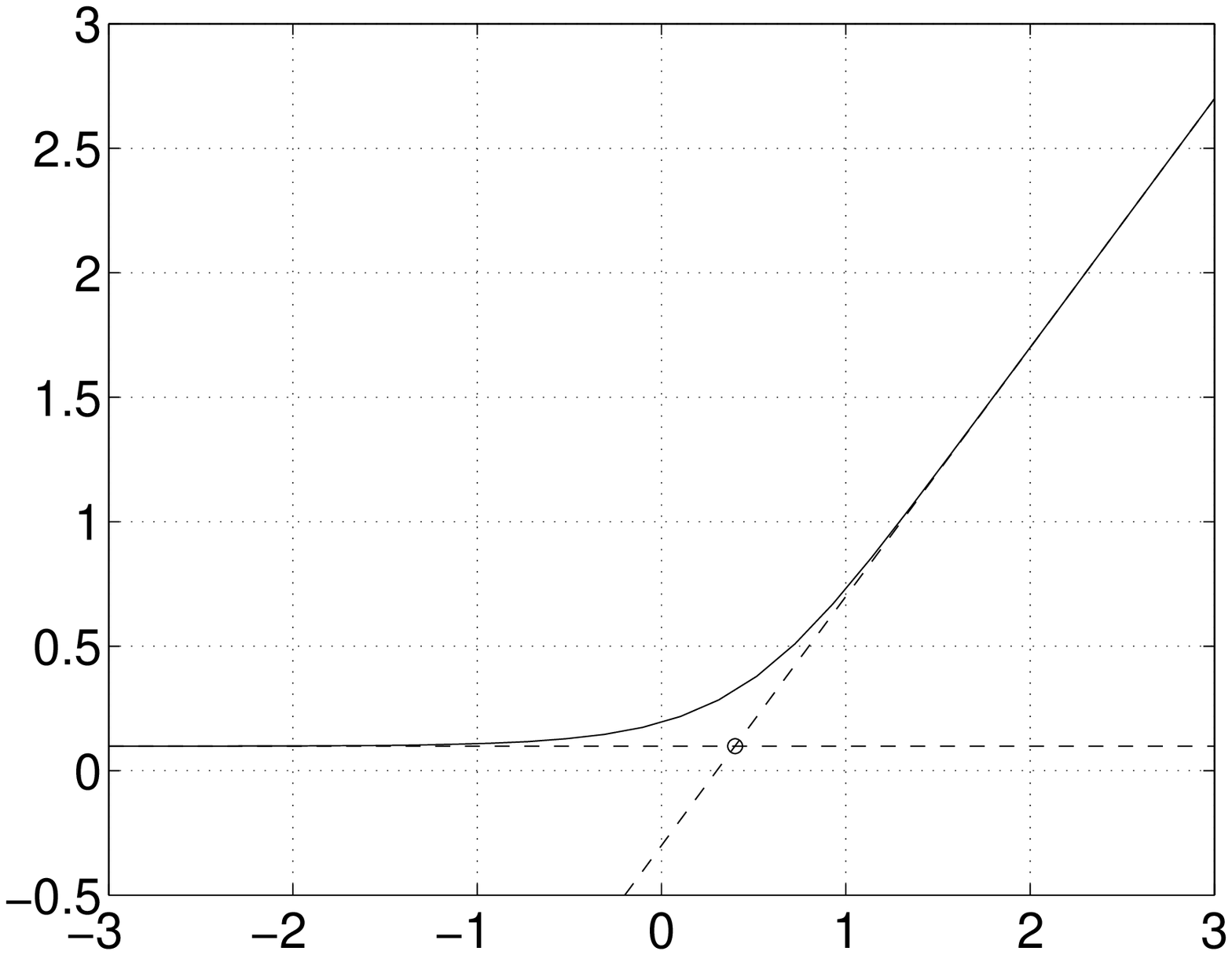}
\ecc
\caption{$\lambda$ and $s$ as a function of $\PAB$, for a fixed $\PDB$ ($\PDB=1$) on a log-log scale. As expected, the plot essentially shows two linear behaviors and a critical case for $\PAB = \sqrt{2\pi}$ ~($\log_{10}\sqrt{2\pi} \sim 0.4$). \label{Fig:LambdaSeuil}}
\end{figure}

\subsection{Laplace Law for Auxiliary Variables} \label{Sec:CasLaplace}

The following developments are dedicated to the case of auxiliary variables under a Laplace law suggested by~\cite{Champagnat04}. 

As mentioned by \cite{Champagnat04} itself, among the Huber-like distributions, such a Laplace-convolved-Gauss probability will have two main advantages: ($i$) the convolution involved in the marginal law  $f_{\Objc}$ (Section~\ref{Sec:Compound}) will be made explicit and ($ii$) the sampling of auxiliary variables  (Section~\ref{Sec:TotalPost}) will be directly feasible thanks to the inversion of the cumulative density function $F_{\Auxc|\Objc}$. 

The Laplace law is written in the form:
\beq \label{Eq:MargAux}
f_{\Auxc}\cro{\Auxb}  = K_{\Auxc}\pmu ~\Exp{ - \, \PAB \, N_1(\Auxb) \, /2} \,,
\eeq
where $\PAB>0$ is a scale parameter, and $N_1(\Auxb) = \sum |\aux_{pq}|$ is the $\mbox{L}_1$ norm. The partition function is simply calculated thanks to separability 
%
%
%
\beqx 
K_{\Auxc} = \displaystyle \int_{\eR^N} \Exp{ - \, \PAB \, N_1(\Auxb) \, /2}  ~\dD \Auxb = \cro{\froc{\PAB}{4}}^{-N}  \,.
\eeqx 
%
%
According to~(\ref{Eq:ObjCondAux}) and~(\ref{Eq:MargAux}) the joint density for $(\Objc,\Auxc)$ takes the form:
\beq \label{Eq:Jointe}
\barr{l} 
~\\
f_{\Objc,\Auxc}\cro{\Objb,\Auxb}  = 
~\\
~\\
~K_{\Objc,\Auxc}\pmu  \Exp{  -\, \PDB \,  N_2( \MatFb \star \Objb - \Auxb) \, + \PAB \, N_1(\Auxb) \, /2} 
~\\
~\\
\earr
\eeq
and the partition function is explicit: $K_{\Objc,\Auxc} = K_{\Objc|\Auxc} \, K_{\Auxc}$.

The marginal law for $\Objc$  involves the one-dimensional convolution of a Gaussian density and a Laplacian density.
\beqx 
\barr{l} 
~\\
~~ f_{\Objc}\cro{\Objb}
~\\
~\\
=\displaystyle K_{\Objc,\Auxc}\pmu  \prod_{pq} \int_{\eR} \Exp{ - \, \PDB \,  \pth{\overline{\obj}_{pq} - \aux_{pq}}^2 \, + \PAB \, |\aux_{pq}| \, /2  } ~\dD \aux_{pq}
~\\
~\\
=\displaystyle K_{\Objc,\Auxc}\pmu  \prod_{pq}  I(+\infty,\overline{\obj}_{pq} , \PDB, \PAB )
~\\
\earr
\eeqx
where $I$ is defined in Appendix~\ref{Sec:AnnL2ConvL1}. Thus, the potential function $\phi$ appears:
\beqx 
f_{\Objc}\cro{\Objb}  = K_{\Objc,\Auxc}\pmu~~ \Exp{ - \,\sum_{pq} \phi(\overline{\obj}_{pq} ) \, /2  } \,,
\eeqx
with: 
\beq \label{Eq:PotentielLogErf}
\phi(x)  = -2 \log I(+\infty, x , \PDB, \PAB ) \,.
\eeq
It is named the Log-Erf potential and it is shown in Fig.~\ref{Fig:Potentiel}. The details of the calculations concerning this potential are given in Appendix~\ref{Sec:AnnLogErf}. Concerning the first derivative, one has:
\beqx
\phi'(0)	= 0  ~~~\AND~~~ \phi'(+\infty)	= \PAB \,,
\eeqx
and concerning the second derivative at origin, one has:
\beqnx
\phi''(0)	&=& \frac{\PAB^2}{2} ~  \cro{ \pth{\eta \, \sqrt{\pi} ~\Erfcx{\eta}}\pmu -1 }
\eeqnx
with  $\eta = \PAB / \sqrt{8\PDB}$ ($\Erfcx{\cdot}$ is given in Appendix~\ref{Sec:AnnErf}). As expected (see Property~\ref{Rem:CVX}), this is a convex potential. It is a $\mbox{L}_2-\mbox{L}_1$ potential which can be reconciled with other more common $\mbox{L}_2-\mbox{L}_1$ potentials (Huber, log-cosh, hyperbolic, fair function). In the case of the Huber potential:
\beq \label{Eq:PotentielHuber}
x \mapsto 
\lambda ~\bca 
x^2								& \IF \vert\, x \,\vert \leq s  \\ 
2s\,\vert\, x \,\vert -s^2	& \IF \vert\, x \,\vert \geq s \\ 
\eca
\eeq
by identifying the second derivatives at zero and the slopes at infinity, one has:
\beq \label{Eq:LambdaSeuil}
\lambda = \frac{1}{2} ~ \phi''(0) ~~~\AND~~~ s = \frac{\phi'(+\infty)}{\phi''(0)} \,.
\eeq
Compared Log-Erf and Huber potentials and their derivatives are shown in Fig.~\ref{Fig:Potentiel}. Using the expansions~(\ref{Eq:ErfcxZero}) and~(\ref{Eq:ErfcxInf}) of Appendix~\ref{Sec:AnnErf}, two limit cases can be identified, according to the value of the ratio  $\eta$:
\beqx
\barr{lrclcrcl}
\FOR~  \eta\gg1 : & \lambda & \simeq & \PDB  										& ; &s &\simeq& \dspsty\frac{\PAB}{2 \, \PDB} \\ ~\\
\FOR~  \eta\ll1 : & \lambda & \simeq & \dspsty\PAB \sqrt{\frac{\PDB}{2\pi}}& ; &s &\simeq& \dspsty\sqrt{\frac{\pi}{2 \, \PDB} }
\earr
\eeqx
In the two limit cases, on a log-log scale, there is linear behavior of $\lambda$ and $s$  as a function of $\PAB$, for a fixed $\PDB$ (see Fig.~\ref{Fig:LambdaSeuil}).  The intersection of the two linear behaviors can be identified as a critical behavior for $\PAB = \sqrt{2\pi\PDB}$. The critical value will be used for the initialization of simulations of Section~\ref{Sec:Simul} (see also Appendix~\ref{Sec:AnnInitHyper}).

\subsection{Practical Case}

In practice, the field is based on a $3\times 3$ Laplacian filter, defined by $[0\,,1\,,0 \,;\, 1\,,-4\,,1 \,;\, 0\,,1\,,0]$ and represented by the matrix $\MatDb$.  At null frequency one has $\rond{\matD}_{\scriptscriptstyle 00}=0$ and as a consequence the mean level of the image is not managed. So, an extra parameter is introduced to drive the mean level: it is denoted by $\eps$ ($\eps\geq0$) and the characteristic matrix $\MatFb$ is set to $\MatFb_{\eps} = \MatDb  + \eps$.

\begin{remark}--- 
If $\eps=0$ the field cannot be normalized and each clique is formed from the four nearest neighbors (cross-like clique). If $\eps\neq0$,  the field can be normalized and each clique is spread out over the entire image.
\end{remark}

The following developments take $\eps>0$ and the partition function of the joint field writes:
\beqx
K_{\Objc,\Auxc}\pmu = \delta ~ \eps  ~ \PDB^{\,N/2} ~ \PAB^{\,N} \,, 
~~~~\WITH~ \delta = \pth{32\pi}^{-N/2}  \prod_{\stackrel{\scriptscriptstyle (p,q)}{\scriptscriptstyle \neq(0,0)}} |\rond{\matD}_{pq}| \,.
\eeqx

Fig.~\ref{Fig:UnPrior} gives a sample of the field with $\PDB=\PAB=1$  and Fig.~\ref{Fig:Histo} gives  histograms of the image pixels, the auxiliary variables $\Aux$  (a Laplace histogram) and the differences $\overline{\Obj}$ (an over-Gaussian histogram).

\begin{remark}---
It is noteworthy that the marginal model $\Objc$ is homogeneous, but the conditional model $\Objc|\Auxc$ is non-homogeneous (except if all the $\aux_{pq}$ are equal). 
\end{remark}

\section{Deconvolution} \label{Sec:Deconvolution}

As a result of the previous Section, a new random field is now available with a special feature: an explicit (and simple) partition function. In the present Section, the field serves as a prior in a deconvolution method whose specificity is to be unsupervised (\ie including hyperparameter estimation).  More precisely, the method relies on a full Bayesian framework and the solution is determined from an \apost law based on an \aprio law (given below) for the object, the noise and the hyperparameters. 




\subsection{Prior choices}\label{Sec:PriorChoice}

\subsubsection{Object law} The \aprio field is defined in the previous section. The joint density for $(\Objc,\Auxc)$  is given by~(\ref{Eq:Jointe}) and it is driven by three parameters: $\PDB$, $\PAB$ and $\eps$. 

\subsubsection{Noise law} The present work is founded on the usual case of zero-mean white Gaussian noise with inverse variance denoted $\PNB$. The density is written:
\beqx
f_{\Bruitc}(\Bruitb) = \pth{2\pi}^{-N/2}~~ \PNB^{N/2} \Exp{ - \, \PNB ~ N_2(\Bruitb) /2 } \,.
\eeqx

\subsubsection{Hyperparameter law} Four parameters are to be managed: $\PNB$, $\PDB$, $\PAB$ and $\eps$. The three parameters of major importance are $\PDNAB=[\PNB, \PDB, \PAB]$; the fourth parameter $\eps$ drives the prior mean level of the image and it is considered as a  nuisance parameter. Anyway, very few is \aprio known about these parameters and the idea is to use non-informative or diffuse and separable priors.

\bit

\item The proposed prior law for the three parameters $\PNB$, $\PDB$ and $\PAB$ is a  conjugate law. It is a gamma law (see Eq.~(\ref{Eq:GammaPDF}), Appendix~\ref{Sec:AnnGamma}) with parameters respectively denoted $(\alpha_\nD,\beta_\nD)$, $(\alpha_\dD,\beta_\dD)$ and $(\alpha_\bD,\beta_\bD)$. It allows for easy computations with the posterior law. Moreover, it includes diffuse and non-informative prior: the uniform prior and the Jeffrey's prior are  obtained as limit cases for $(\alpha,\beta)=(1,\infty)$ and for $(\alpha,\beta)=(0,\infty)$ respectively. 

\item The last parameter $\eps$ is considered as a  nuisance parameter and the proposed strategy resorts to integration out. The desired prior law is a Dirac law,  so that no information is accounted for about the mean level of the image (it is set on the basis of observed data only). Formally, in a first step a uniform density over $[\,0,M_\eps\,]$ is introduced  and in a second step the limit law for $M_\eps\rightarrow 0$ is considered.

\eit


\subsection{Joint Law} \label{Sec:PostJoint}

Thus, the joint law is established for $(\Datac,\Objc,\Auxc,\Cc,\Ec)$:
\beqx
\barr{l}
f_{\Datac,\Objc,\Auxc,\Cc,\Ec}(\Datab,\Objb,\Auxb,\PDNAB,\eps) ~ =    ~\\ ~\\
~~~ \delta' ~ \PNB^{~\alpha_\nD-1+N/2} ~ \PDB^{~\alpha_\dD-1+N/2} ~ \PAB^{~\alpha_\bD-1+N} ~ \eps ~ M_\eps\pmu \unbb_{[\,0,M_\eps\,]}(\eps)\\ ~\\ 
~~  \exp - \, \acc{ Q_\eps /2 ~+~ \PNB/{\beta_\nD} ~+~ \PDB/{\beta_\dD} ~+~ \PAB/{\beta_\bD} } ~\\~\\
\earr
\eeqx
where $\delta'= \pth{2\pi}^{-N/2} \delta  / \beta_\nD^{\alpha_\nD} \, \Gamma[\alpha_\nD] ~ \beta_\bD^{\alpha_\bD} \, \Gamma[\alpha_\bD] ~ \beta_\dD^{\alpha_\dD} \, \Gamma[\alpha_\dD]$ is a normalization constant 
and $Q_\eps$ is part of the Co-logarithm of the density involving $\MatFb_{\eps}$: 
\beqx
Q_\eps  = \PNB \, N_2(\Datab - \MatHb \star \Objb)	 + \PDB \, N_2(\MatFb_{\eps} \star \Objb - \Auxb) +  \PAB \, N_1(\Auxb) \,.
\eeqx
The \apost density is formed for $\Objc$, $\Auxc$, $\Cc$ and $\Ec$, given $\Datac$ thanks to the Bayes rule:
\beqx
\barr{l}
f_{\Objc,\Auxc,\Cc,\Ec | \Datac}(\Objb,\Auxb,\PDNAB,\eps | \Datab) =  ~\\ ~\\ 
~~~~~~		\dspsty \frac{f_{\Objc,\Auxc,\Cc,\Ec,\Datac}(\Objb,\Auxb,\PDNAB,\eps,\Datab) }{\dspsty \int_{\Objb,\Auxb,\PDNAB,\eps} f_{\Objc,\Auxc,\Cc,\Ec,\Datac}(\Objb,\Auxb,\PDNAB,\eps,\Datab) \, \dD\Objb  \, \dD\Auxb  \, \dD\PDNAB  \, \dD\eps  } \,,
\earr
\eeqx
and it is parametrized by the $(\alpha,\beta)$ and $M_\eps$. Then, $\eps$ is integrated out and the law for $\Objc$, $\Auxc$, $\Cc$  given $\Datac$ writes
\beqx
f_{\Objc,\Auxc,\Cc  | \Datac}(\Objb,\Auxb,\PDNAB | \Datab) = \dspsty \int_{\eps} f_{\Objc,\Auxc,\Cc,\Ec   | \Datac}(\Objb,\Auxb,\PDNAB,\eps  | \Datab) \, \dD\eps  \,.
\eeqx
It is also parametrized by the $(\alpha,\beta)$ and $M_\eps$, so, the limit is set when $M_\eps$ tends to 0. The detail of the calculations is given in Appendix~\ref{Sec:AnnIntegreHyper} and it is shown that a probability density function is obtained if the mean level of the object is observed, \ie $\rond{\matH}_{\scriptscriptstyle 00}\neq0$.

\subsection{Posterior Law and Posterior Mean}\label{Sec:TotalPost}\label{Sec:PosteriorMean}

Thus, the Total Posterior Law can be deduced for all the unknown parameters $(\Objc,\Auxc,\Cc)$ given the observed data $\Datac$:
\beq \label{Eq:TotalPosterior}
\barr{l}
f_{\Objc,\Auxc,\Cc|\Datac}(\Objb,\Auxb,\PDNAB|\Datab) ~ \propto  ~\\ ~\\
~~~~~~ \PNB^{~\alpha_\nD-1+N/2} ~~ \PDB^{~\alpha_\dD-1+N/2} ~~ \PAB^{~\alpha_\bD-1+N} ~\\ ~\\ 
~~~~~  \exp - \, \acc{ Q_0 /2 ~+~ \PNB/{\beta_\nD} ~+~ \PDB/{\beta_\dD} ~+~ \PAB/{\beta_\bD} } ~\\~\\
\earr
\eeq
where $Q_0$ involves $\MatFb_{0}=\MatDb$: 
\beqx
Q_0  = \PNB \, N_2(\Datab - \MatHb \star \Objb)	 + \PDB \, N_2(\MatDb \star \Objb - \Auxb) +  \PAB \, N_1(\Auxb) \,.
\eeqx


In practice, the chosen point estimate is the posterior mean (\ie the Minimum Mean Square Error). Its calculation is performed by means of Monte-Carlo Markov Chain stochastic sampling algorithm~\cite{Robert04,Winkler03}: auxiliary variables, object and hyperparameters are successively sampled given the other in a Gibbs strategy.

\subsubsection{Sampling auxiliary variables}\label{Sec:SampleAux} The sampling of auxiliary variables is delicate but can be directly done. It is based on the inversion of the cumulative density function (cdf) $F_{\Auxc|\Objc}$. It is sufficient to uniformly sample $u$  in $[0,1]$ and to compute $\aux=F_{\Auxc|\Objc}\pmu(u)$. The calculations can be found in Appendix~\ref{Sec:AnnInvCDF}.

\subsubsection{Sampling object}\label{Sec:SampleObject} The object is a  toroidal Gaussian field and the $\rond{\Obj}_{pq}$  are independent with  mean $\rond{\mu}_{pq}$ and inverse variance $\rond{\nu}_{pq}$ (see calculations in Appendix~\ref{Sec:AnnObjCondPost})
\beqn
\rond{\nu}_{pq} &=& \PNB \, |\rond{\matH}_{pq}|^2 + \PDB \, |\rond{\matD}_{pq}|^2 \label{Eq:MuNuNU}\\ 
\nonumber~\\
\rond{\mu}_{pq} &=& \cro{\,\PNB \, \rond{\matH}_{pq}^{\,*} ~ \rond{\data}_{pq} + \PDB \, \rond{\matD}_{pq}^{\,*} ~ \rond{\aux}_{pq}} / \, \rond{\nu}_{pq}\label{Eq:MuNuMU}
\eeqn  
where superscript $~^*$ stands for the complex conjugate. Thus, the sampling is reduced to the sampling of an inhomogeneous white Gaussian noise followed by an \fftdd.

\begin{table*}
\bcc
\begin{tabular}{lcccccccc} 
						& Data		& PM			& CPM			& CPM 				& CPM					& CPM 			&MAP-LogErf	& MAP-Huber	\\ 
						&				&				&				& (best $\PAB$)	& (best $\PDB$)	& (best $\PNB$)&				&				\\
\hline\hline
Dist. $\mbox{L2}$	& 11.62\%	& 3.93\%		& 3.94\%		& 3.87\%				& 3.85\%				& 3.81\%			& 5.56\%		& 5.63\%		\\ \hline
Dist. $\mbox{L1}$	& 35.47\%	& 19.47\%	& 19.50\%	& 18.92\%			& 19.40\%			& 19.18\%		& 20.68\%	& 20.84\%	\\ \hline
\\
\end{tabular}
\ecc
\caption{Quantitative comparison by means of $\mbox{L2}$ and $\mbox{L1}$ distances between true image and data (column 1), true image and estimated  images (column 2 to 8). \label{Tab:ErrMesures}}
\end{table*}

\begin{figure*}[htb]
\bcc\btabu{rrrr}
\includegraphics[width=4cm]{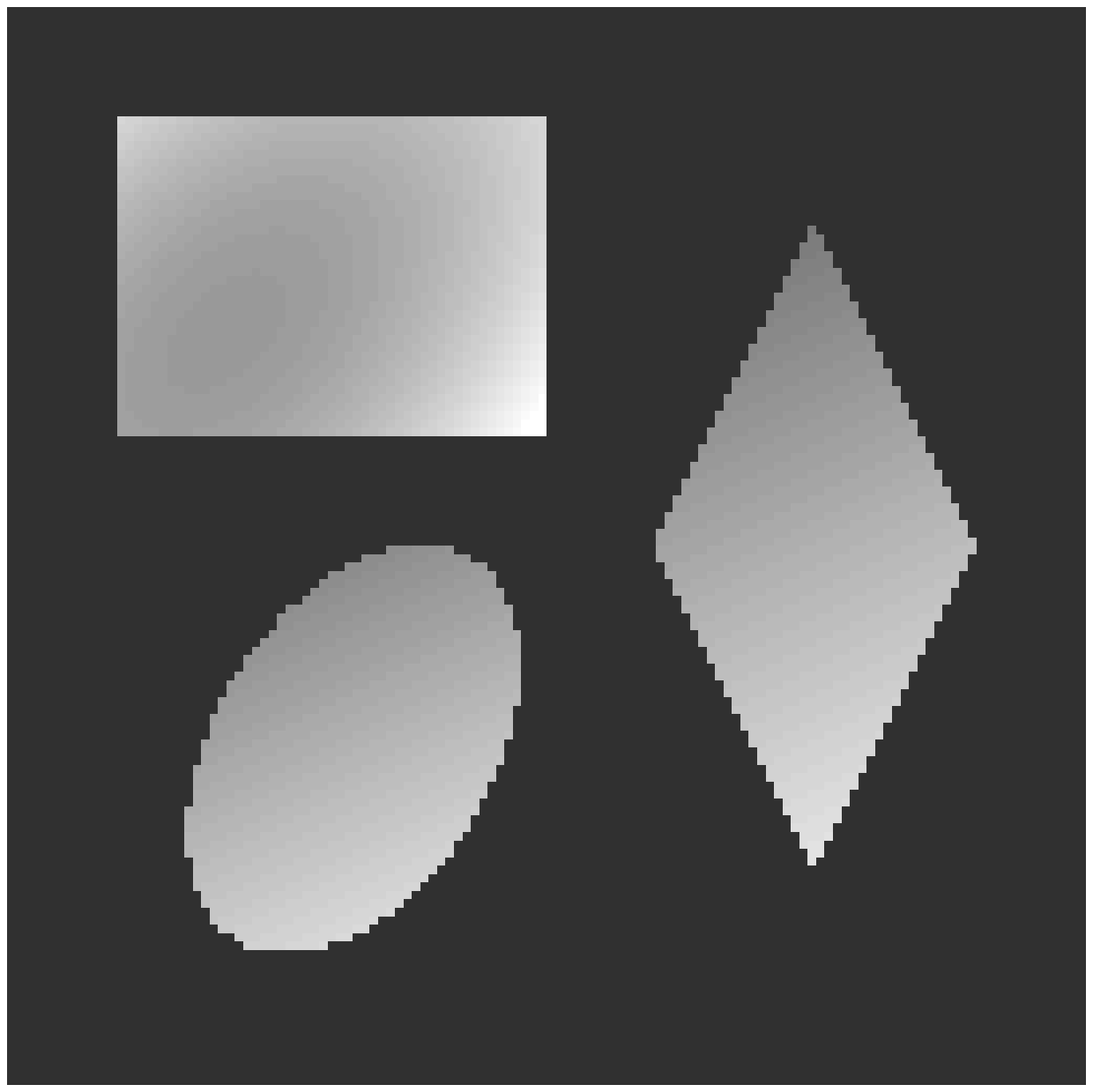}&
\includegraphics[width=4cm]{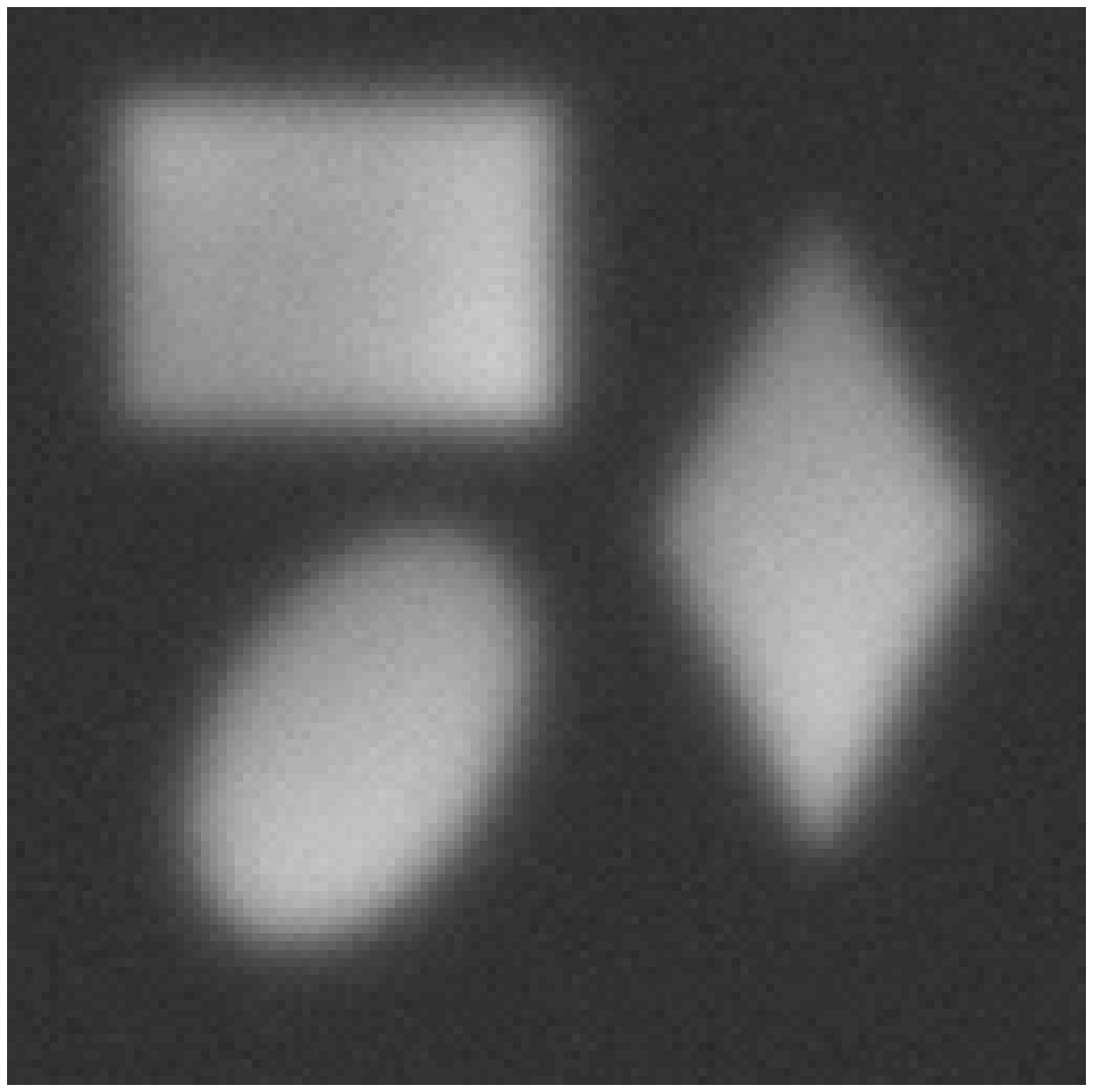}&
\includegraphics[width=4cm]{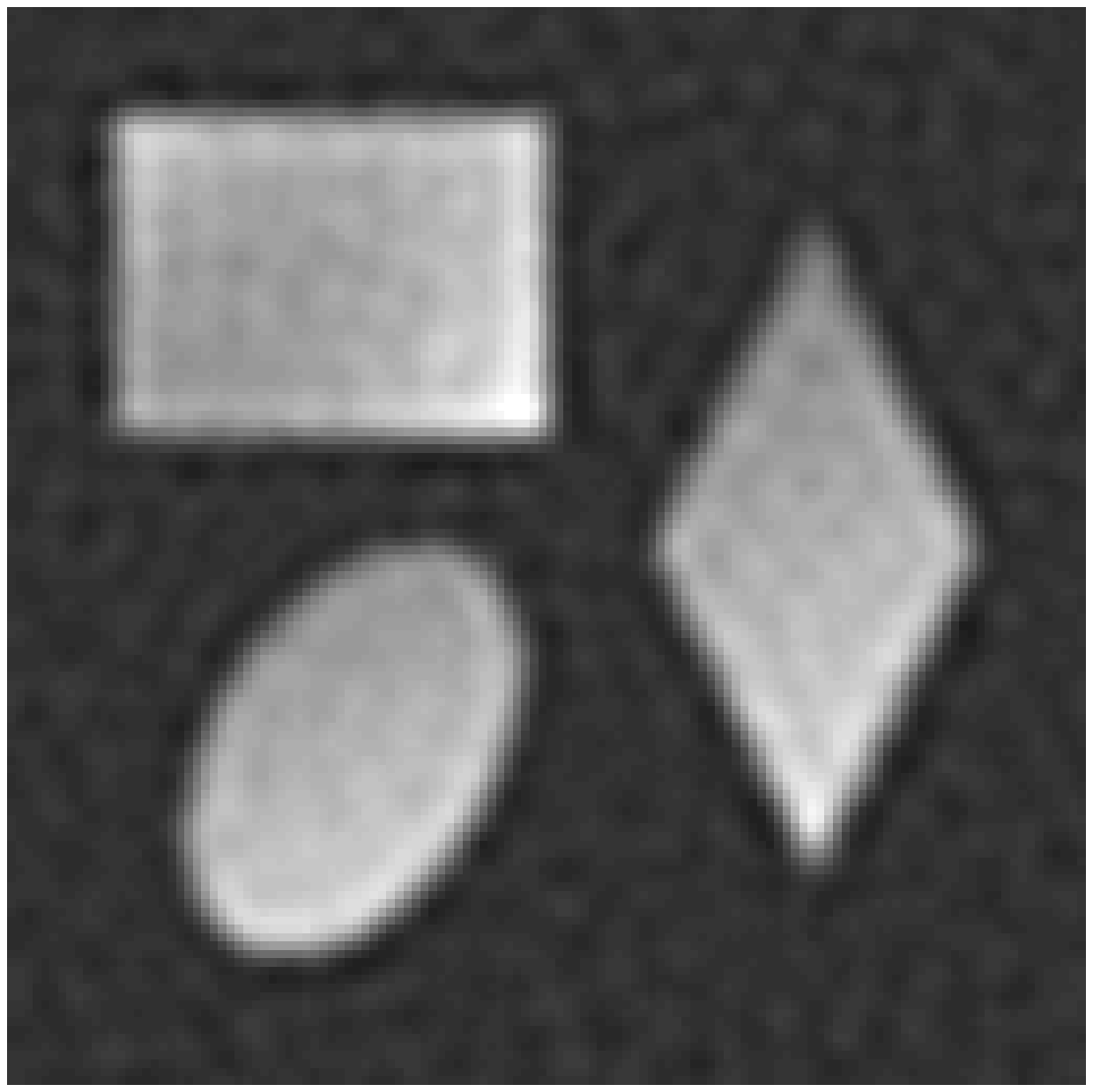}&
\includegraphics[width=4cm]{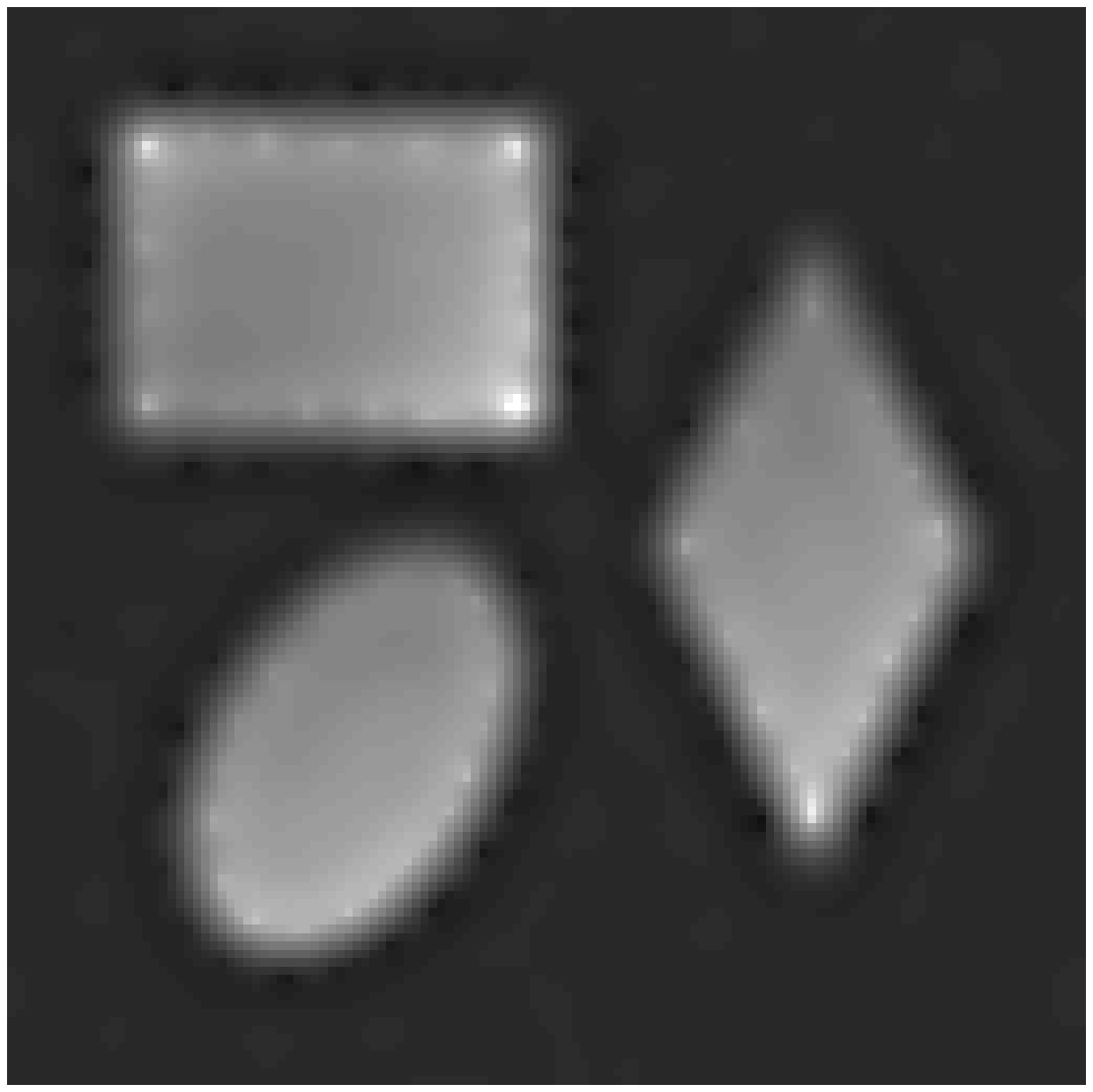}\\
\includegraphics[width=4.1cm]{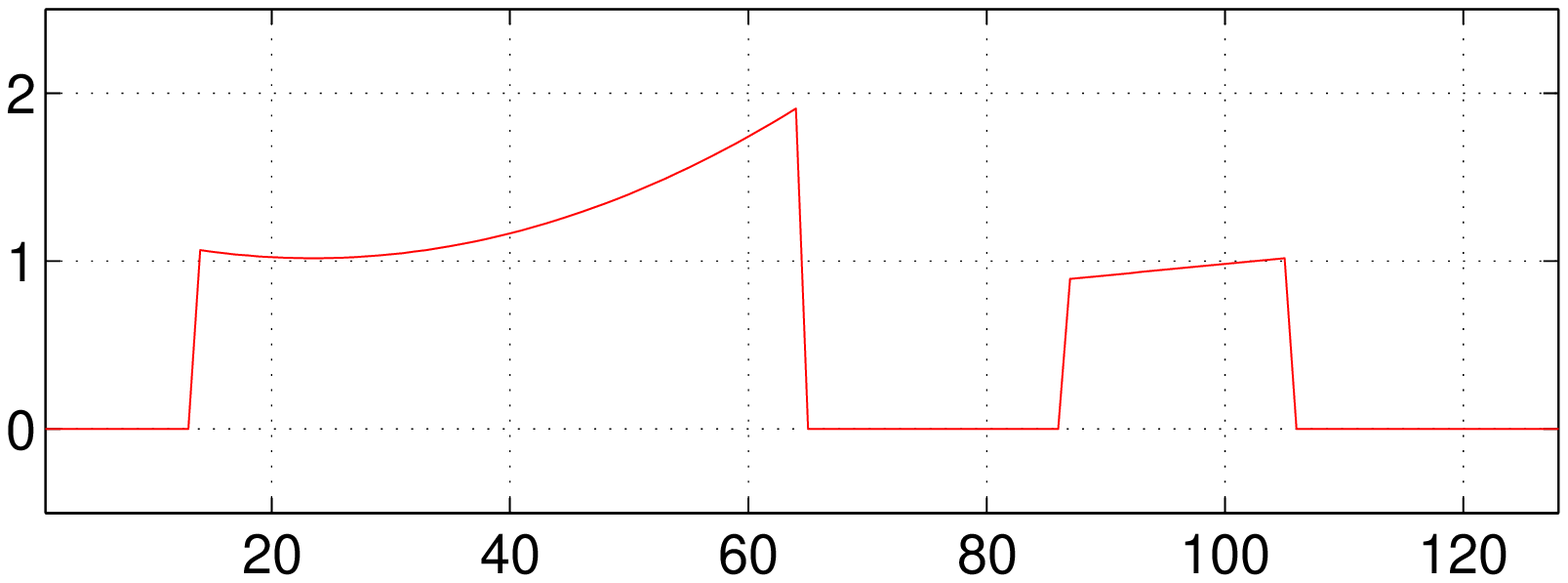}&
\includegraphics[width=4.1cm]{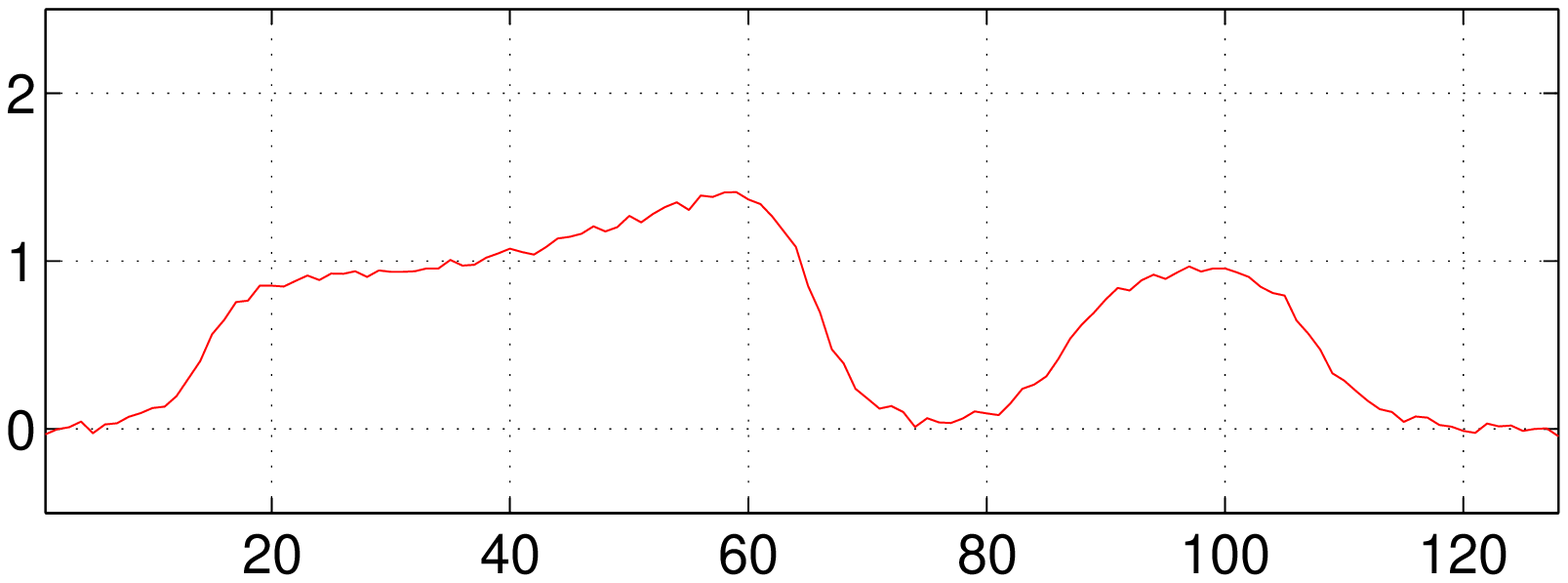}&
\includegraphics[width=4.1cm]{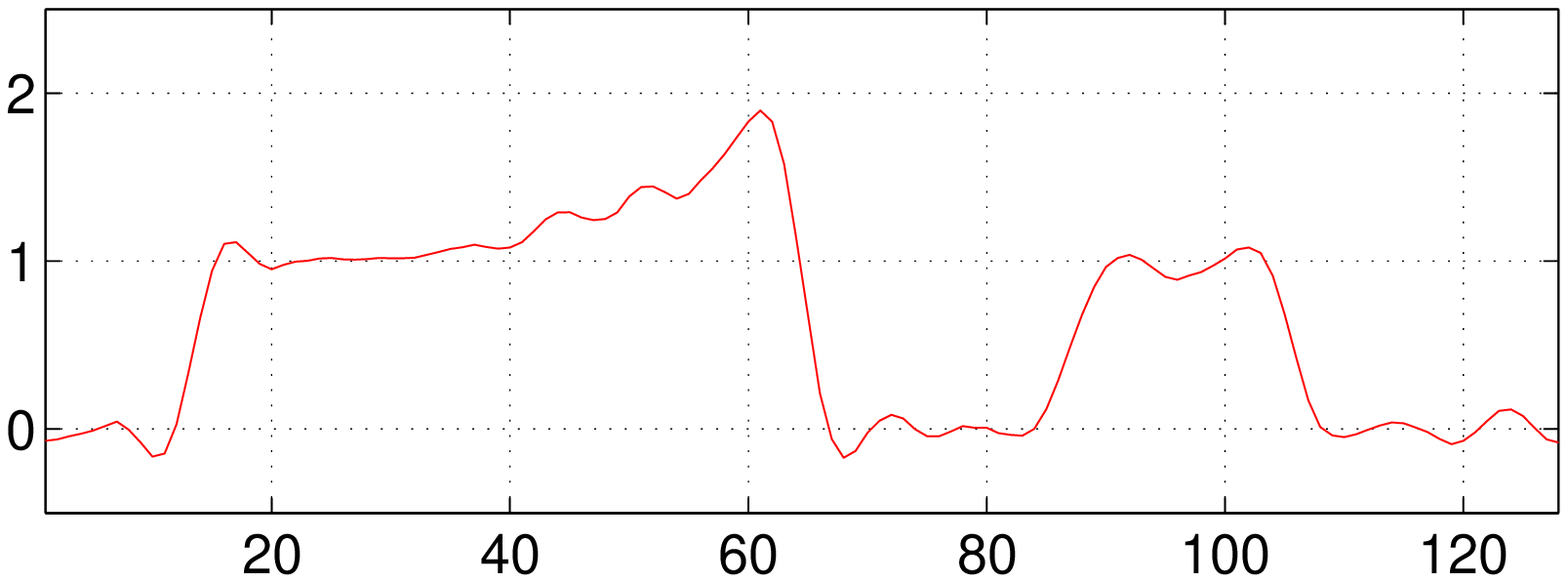}&
\includegraphics[width=4.1cm]{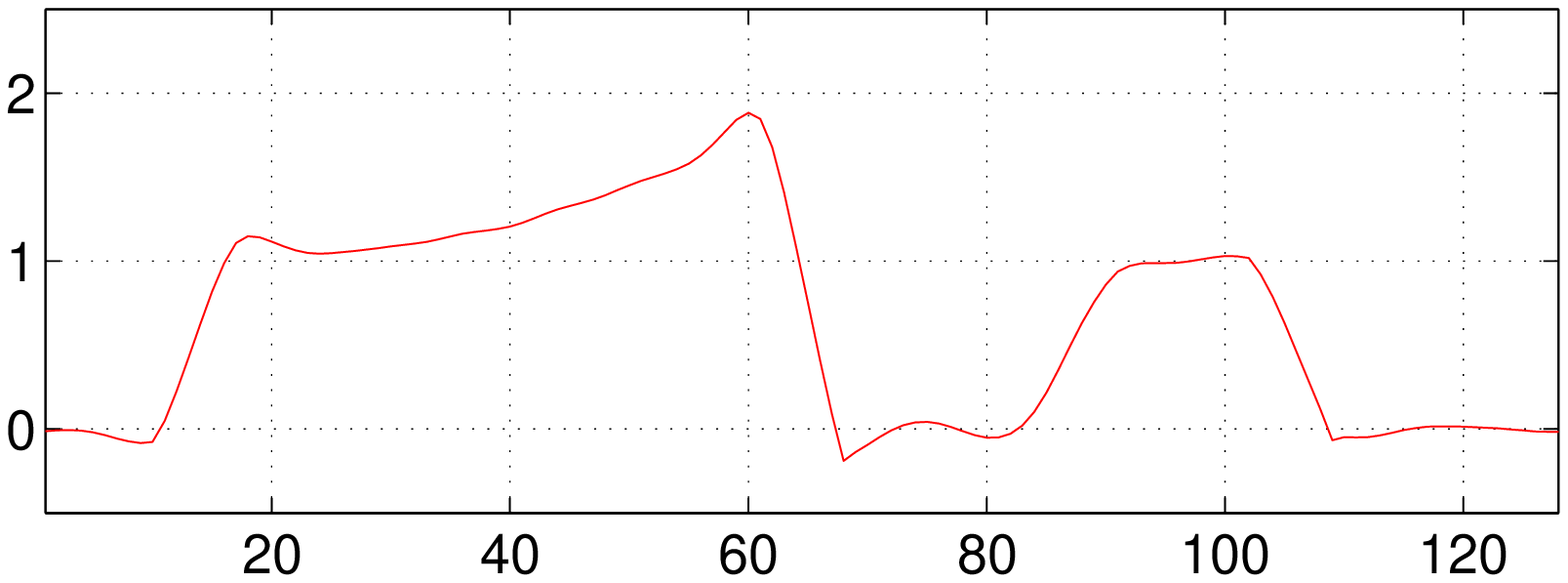} \\
\etabu\ecc
\caption{From left to right : original image $\Objb^\star$, observed data $\Datab$, deconvolved image $\wh{\Objb}\PostMean$ and deconvolved image $\wh{\Objb}\MAP$. At the top: gray level images and at the bottom: profile of the 100-th row (which encroaches on both the rectangle and the rhombus). In order to evaluate the relative dynamics in each case, all the images are shown in the same gray-scale between -0.5 and 2. The four shown profiles are also presented between -0.5 and 2. \label{Fig:Resultats}}
\end{figure*}

\subsubsection{Sampling hyperparameters}\label{Sec:SampleHyper}

Each parameter $\PNB$, $\PDB$ and $\PAB$ follows a gamma\footnote{The sampling of the Gamma variables is achieved using the Matlab function \texttt{gamrnd}.} law derived form~(\ref{Eq:TotalPosterior}) (see Appendix~\ref{Sec:AnnGamma}) with respective parameters $\alpha$ and $\beta$ 
\beqx
\barr{lcl}
\alpha = \alpha_\nD + N/2 &\AND& \beta\pmu = \beta_\nD\pmu + \froc{N_2(\Datab - \MatHb \star \Objb )}{2} \\
\alpha = \alpha_\dD + N/2 &\AND& \beta\pmu = \beta_\dD\pmu + \froc{N_2(\MatDb \star \Objb - \Auxb)}{2} \\
\alpha = \alpha_\bD + N   &\AND& \beta\pmu = \beta_\bD\pmu + \froc{N_1(\Auxb)}{2} \,.
\earr
\eeqx

The description of the method and the algorithm are now complete and synthesized in Table~\ref{Tab:Algo}. The remainder of this Section illustrates the implementation practicability. 

\begin{table}
\begin{fminipage}
\smallskip

\texttt{Initialize}

	\bit
	\item \texttt{N = 1, Delta = inf}
	\item \texttt{HatX = Data}
	\item \texttt{GamN, GamD, GamB } (Annex~\ref{Sec:AnnInitHyper})
	\eit

	\bigskip

\texttt{Repeat}

	\bit
	\item \texttt{Sample step}
	
		\bit
		\item \texttt{Auxiliary variables B} (Sect. \ref{Sec:SampleAux})
		\item \texttt{Object X} (Sect. \ref{Sec:SampleObject})
		\item \texttt{Hyperparameter GamN,GamD,GamB} (Sect.\,\ref{Sec:SampleHyper})
		\eit

	\medskip
	\item \texttt{Update}
	
		\bit
		\item \texttt{N = N+1}
		\item \texttt{Delta = ( HatX - X ) / N}
		\item \texttt{HatX = ( (N-1)*HatX + 1*X ) / N}
		\eit

	\eit

\texttt{Until Delta<eps}

\smallskip
\end{fminipage}
\caption{Detailed algorithm (pseudo-code). \label{Tab:Algo}}
\end{table}

\subsection{Computation feasibility} \label{Sec:Simul}


This part illustrates the previous developments and it only aims at demonstrating the numerical practicability of the method. It is built on a deliberately simple image $\Objb^\star$ appropriate in order to evaluate the capabilities and the limitations of the proposed approach: the image is set up from homogeneous zones separated by sharp edges (see Fig.~\ref{Fig:Resultats}, on the left). It is a $128\times128$ image composed of a black background and three objects with gray levels gradually changing between 0.7 and 2.1. The difference between neighboring pixels varies between 0 and 2.1 in absolute value. Regarding the Laplacian of the image, $\overline{\Objb} = \MatFb \star \Objb$, the set of $\overline{\Obj}_k$ can be split in two sets: 94\,\% of the $\overline{\Obj}_k$ are less than $2.10^{-4}$ (inside homogeneous zones) and 6\,\% of the $\overline{\Obj}_k$ are greater than $3.10^{-2}$ (located around edges). No value is between $2.10^{-4}$ and  $3.10^{-2}$. 

The impulse response of the system is Gaussian shaped with 6 pixels width at half-maximum, the noise variance is $2.10^{-2}$ and the resulting observed image $\Datab$ is shown in Fig.~\ref{Fig:Resultats} (in the second column). The resolution is clearly degraded and details of the edges are no longer visible (neither on the gray level image nor on the shown profile). The dynamic is also strongly affected, notably at about the 64-th sample of the shown profile.


The procedure is initialized by the empirical least-squares hyperparameters given in Appendix~\ref{Sec:AnnInitHyper}. The object $\Objb$ is initialized by the observed data (and there is no need to initialize the auxiliary variables). Moreover, practically, the $(\alpha_*,\beta_*)$  are set to $(0,\infty)$ corresponding to the Jeffrey's prior.

The proposed algorithm\footnote{The proposed algorithm has been implemented with the computing environment Matlab on a PC, with a 2~GHz AMD-Athlon CPU, and 512~MB of RAM. Code is $\sim$ 100 lines long.} generates samples of the \apost law $f_{\Objc,\Auxc,\Cc|\Datac}(\Objb,\Auxb,\PDNAB|\Datab)$. Practically, the algorithm behaves as expected: the stationary law is attained after a burn-in time (about 200 iterations) and remains in a steady behavior. The empirical mean of the generated images is recursively computed and the algorithm is stopped when its variation becomes smaller than a given value $T$ (in quadratic norm). In the presented example $T=5.10^{-4}$, the algorithm produced 953 iterations and computation time was 47 seconds. 


%

\begin{figure*}[htb]
\bcc
\btabu{ccc}
\includegraphics[height=4cm]{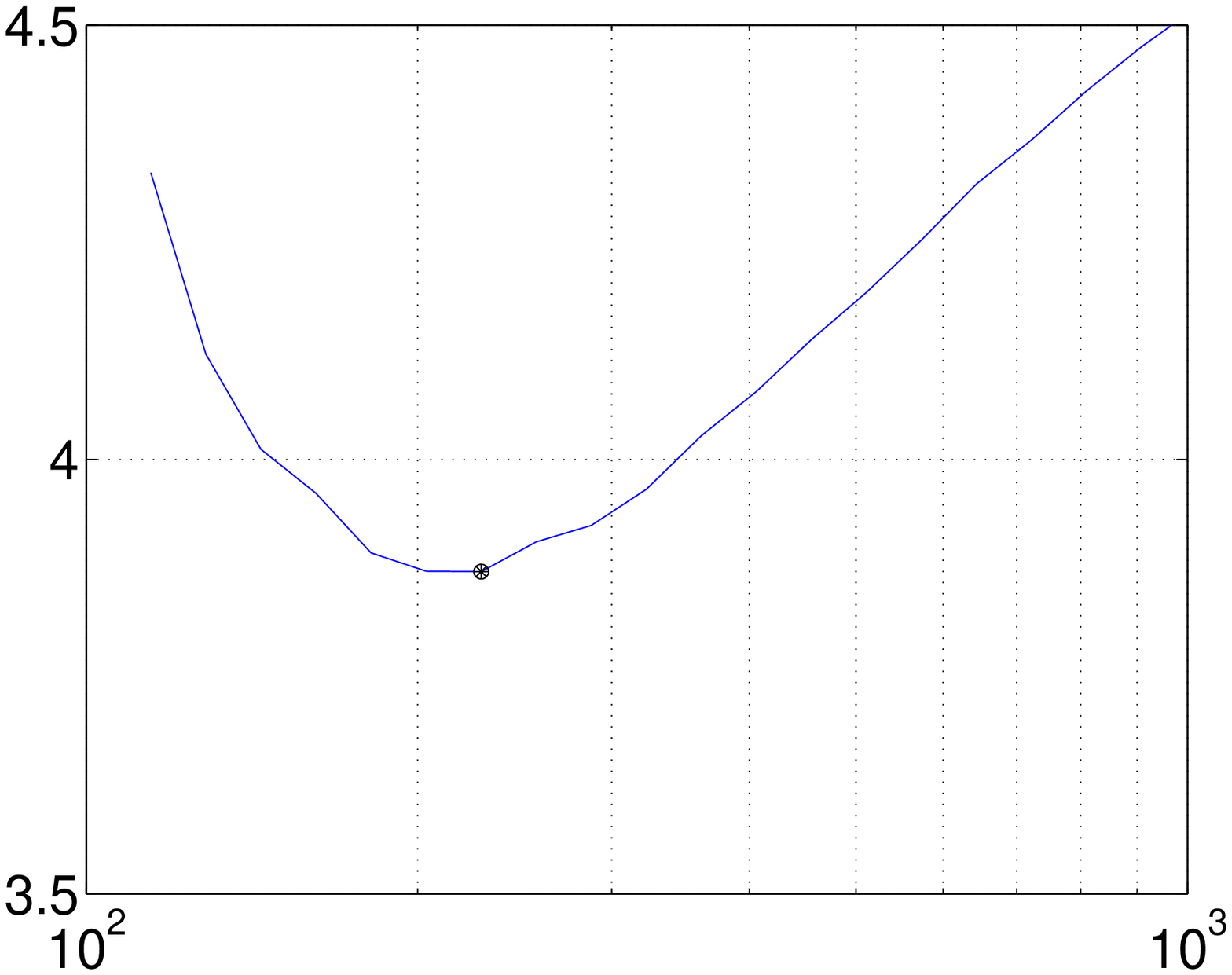} &\includegraphics[height=4cm]{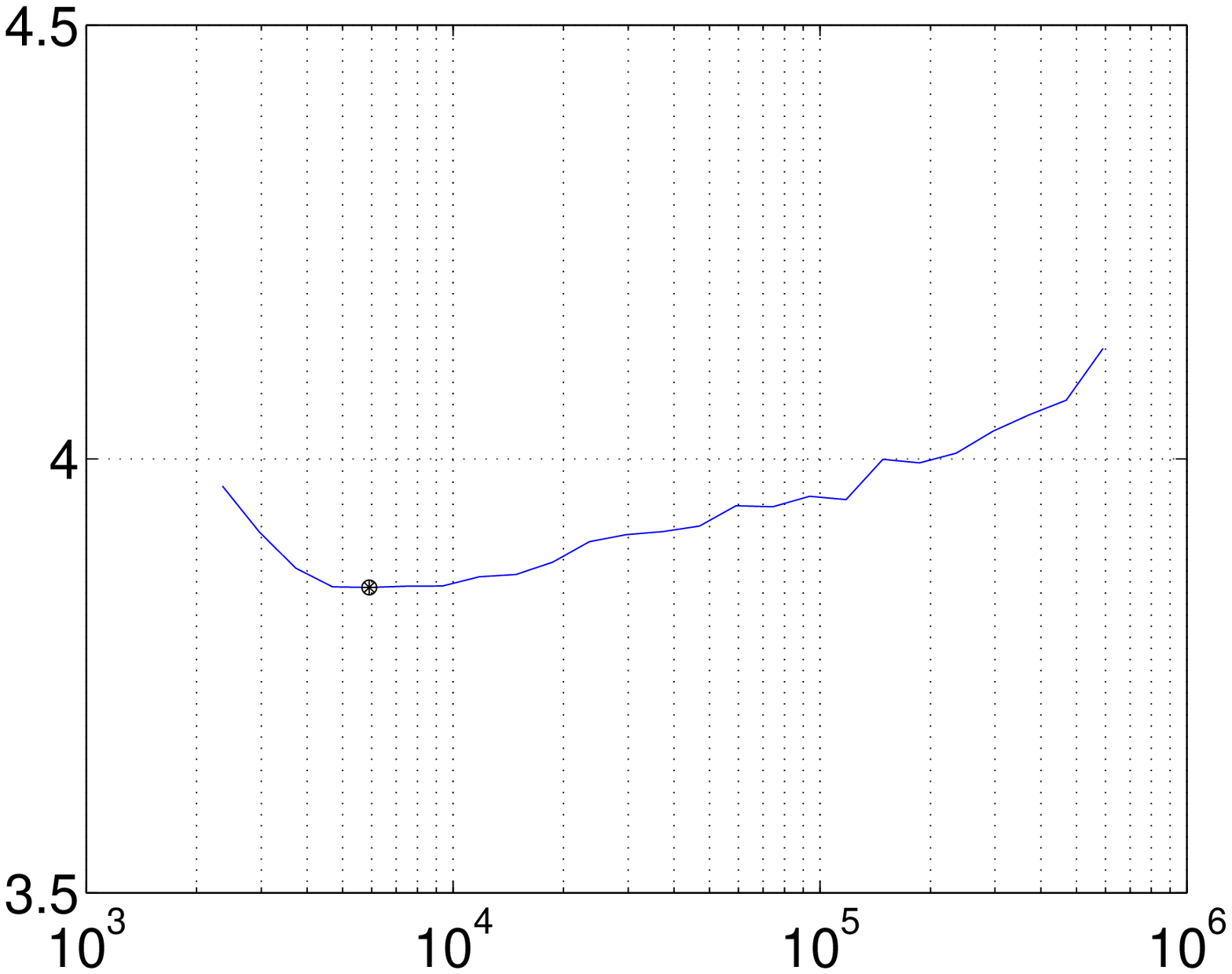} &\includegraphics[height=4cm]{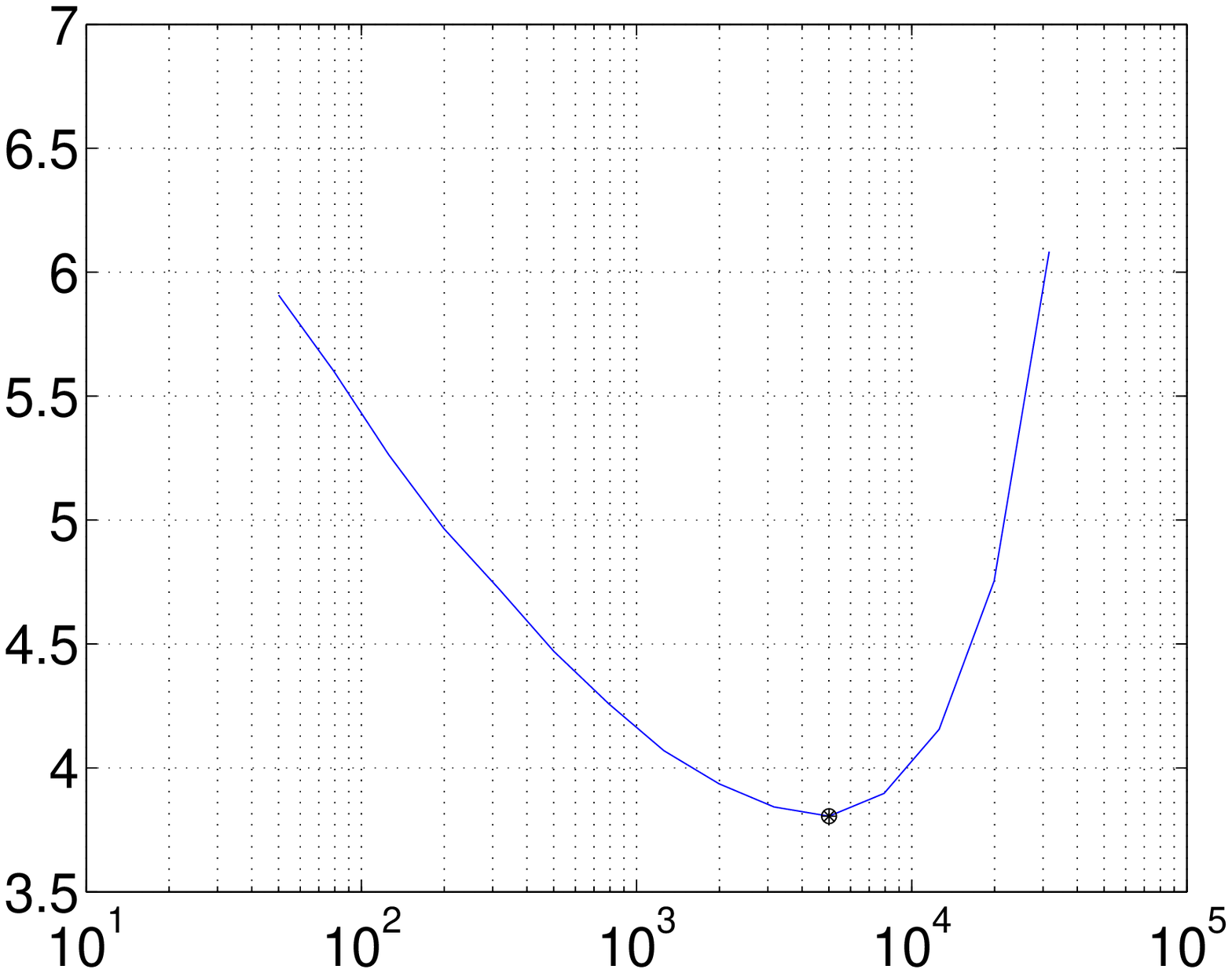} \\
\includegraphics[height=4cm]{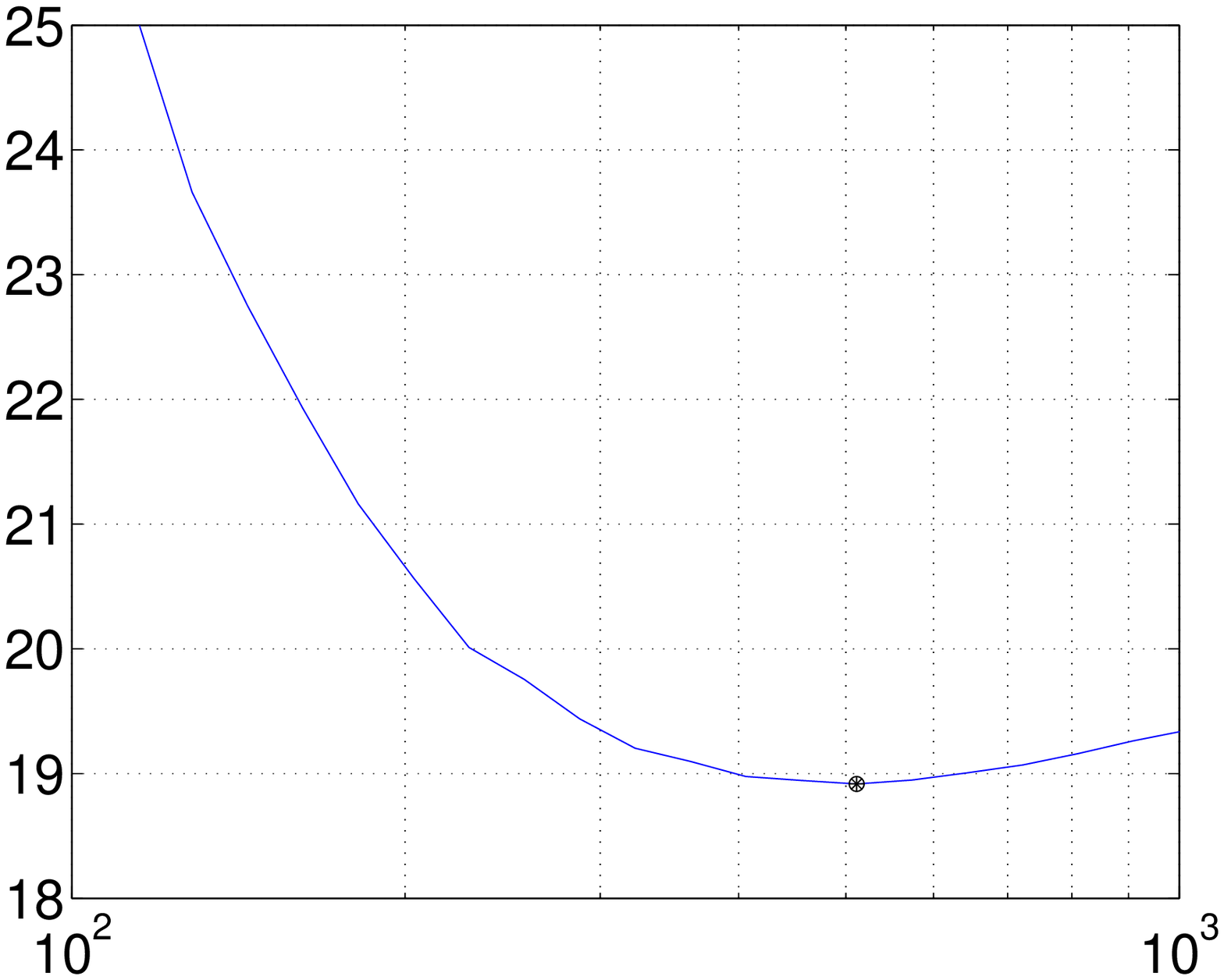}&\includegraphics[height=4cm]{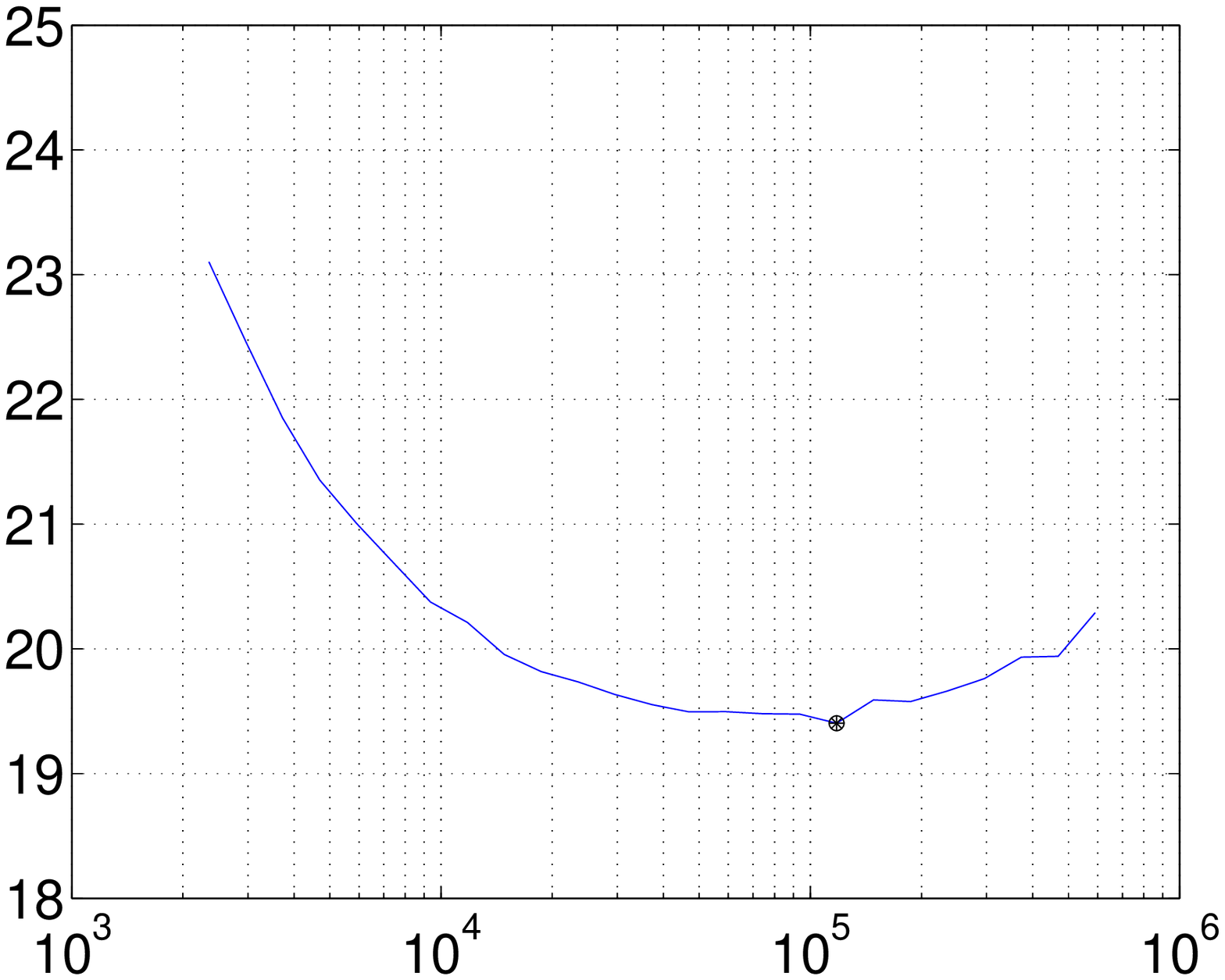}&\includegraphics[height=4cm]{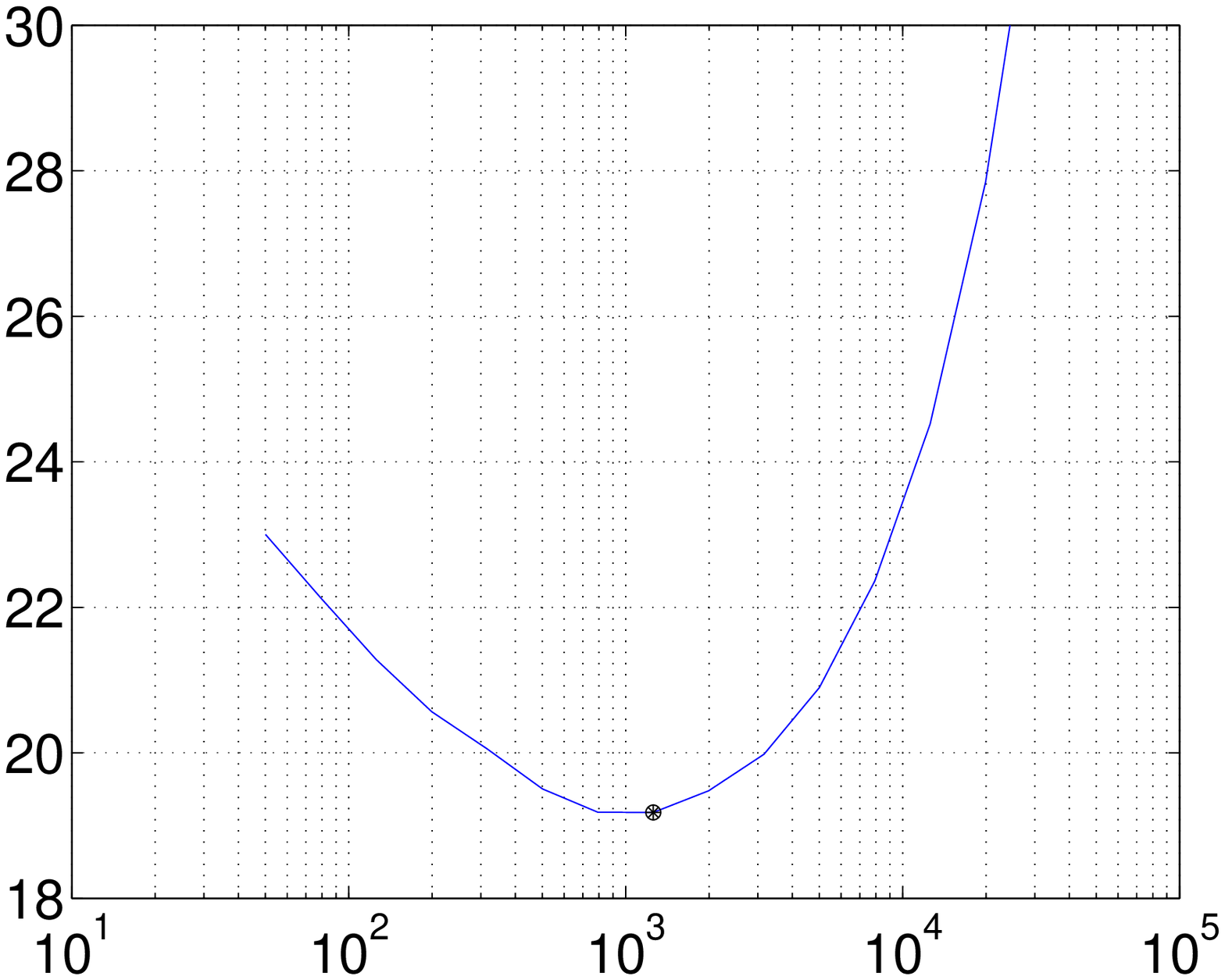} \\
\etabu
\ecc
\caption{Distances between the true image $\Objb^\star$ and Conditional Posterior Mean $\wh{\Objb}\CondPostMean$ as a function of the parameters $\PAB$, $\PDB$ and $\PNB$, around the \post mean value $\wh{\PDNAB}$. From left to right: error is shown as a function of $\PAB$, $\PDB$ and $\PNB$. Top row gives L2 distance and bottom row gives L1 distances. The black dots give the minimum distances reported in Table~\ref{Tab:ErrMesures}. \label{Fig:MesureErreur}}
\end{figure*}

\begin{figure}[htb]
\bcc\btabu{rl}
\includegraphics[height=3.5cm]{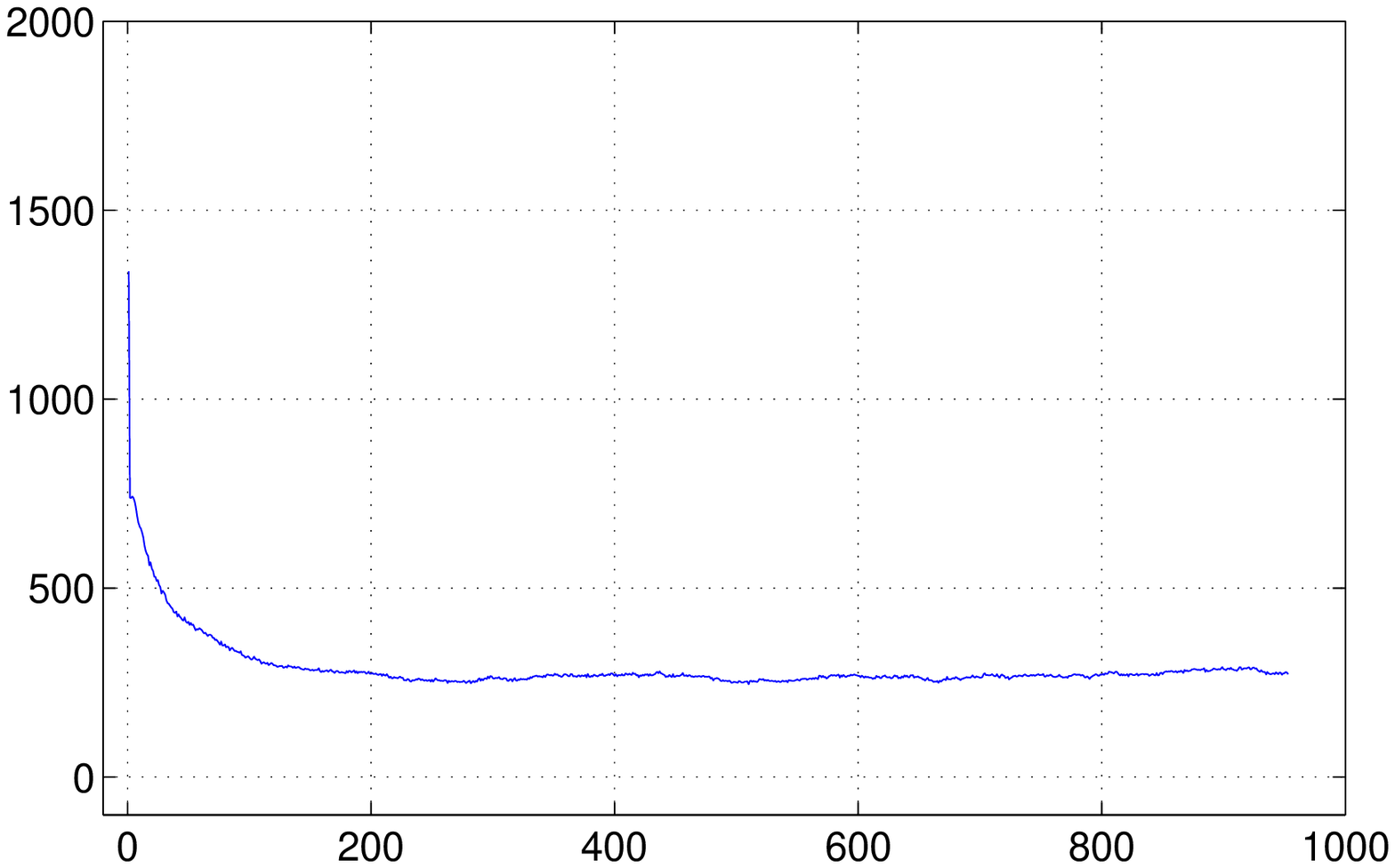} & \includegraphics[height=3.5cm]{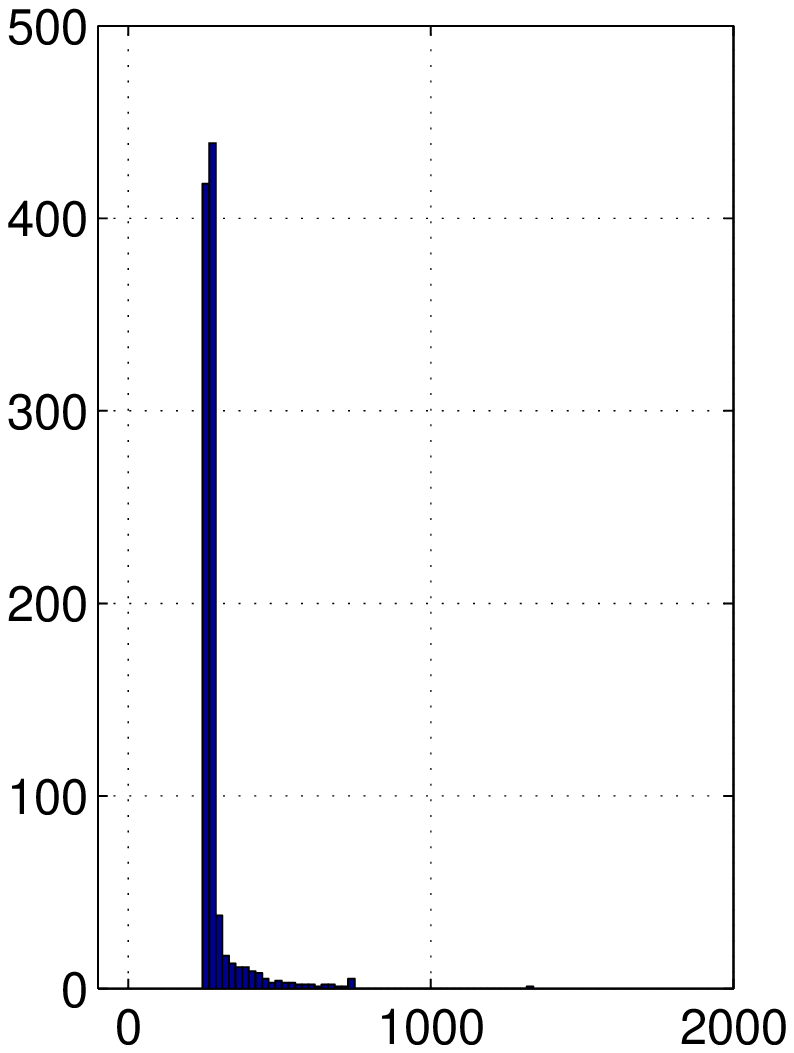} \\
\includegraphics[height=3.55cm]{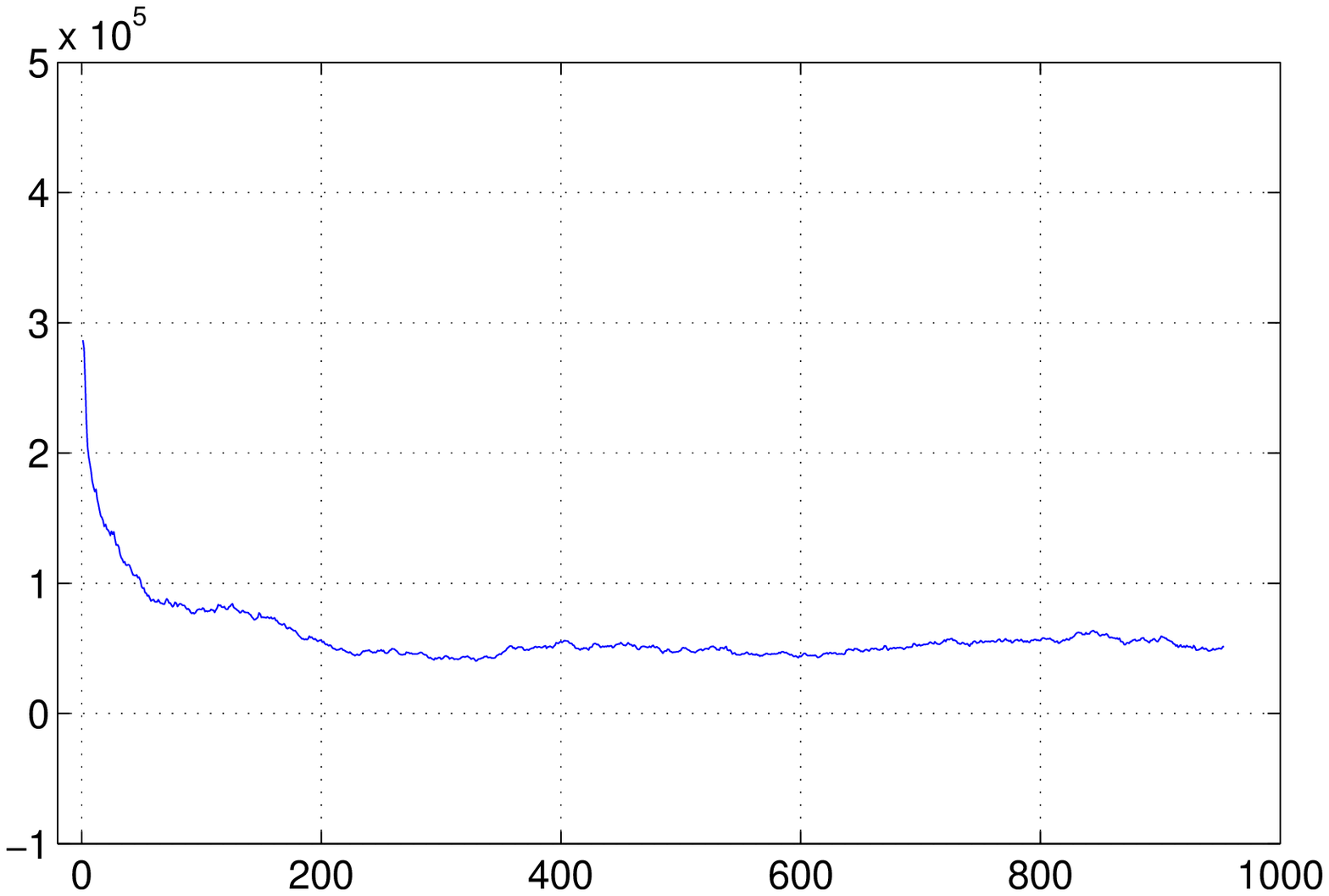} & \includegraphics[height=3.4cm]{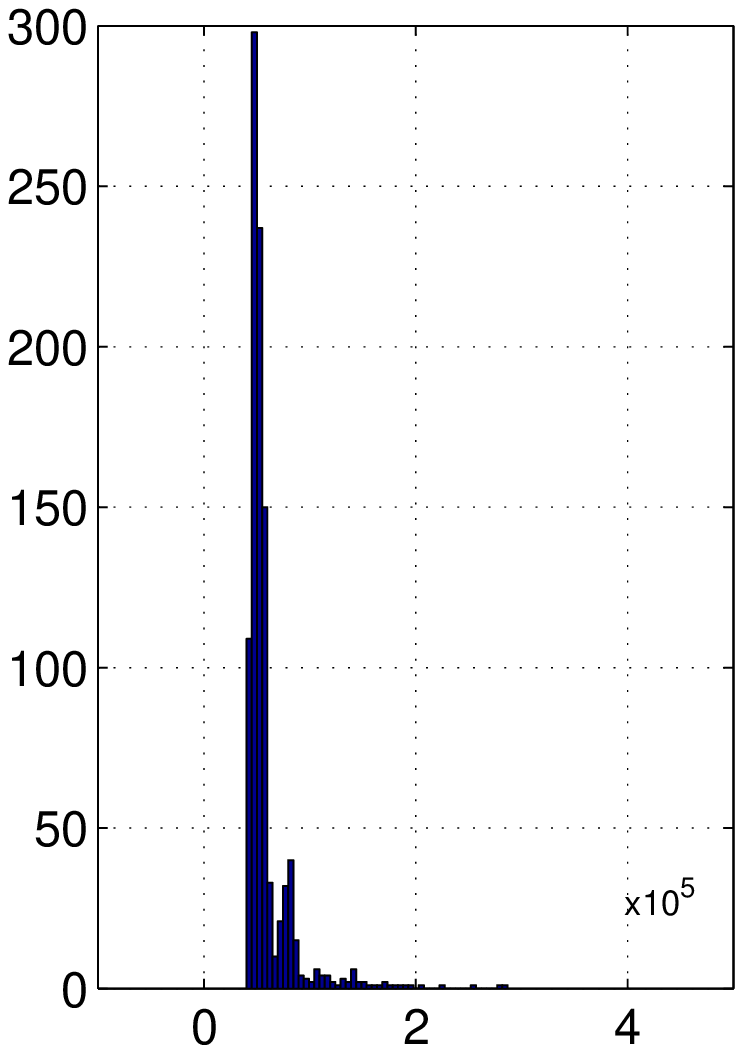} \\
\includegraphics[height=3.5cm]{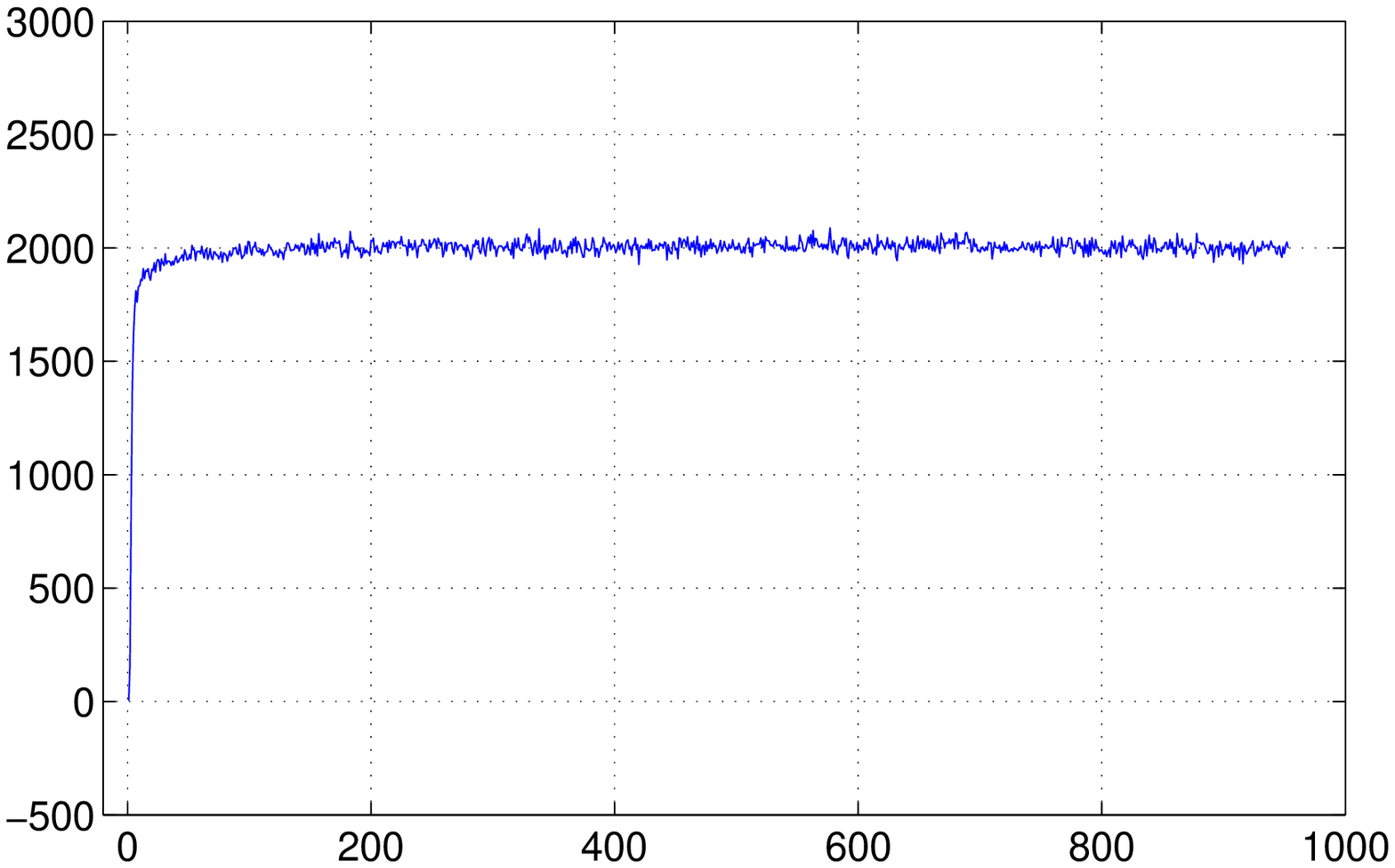} & \includegraphics[height=3.5cm]{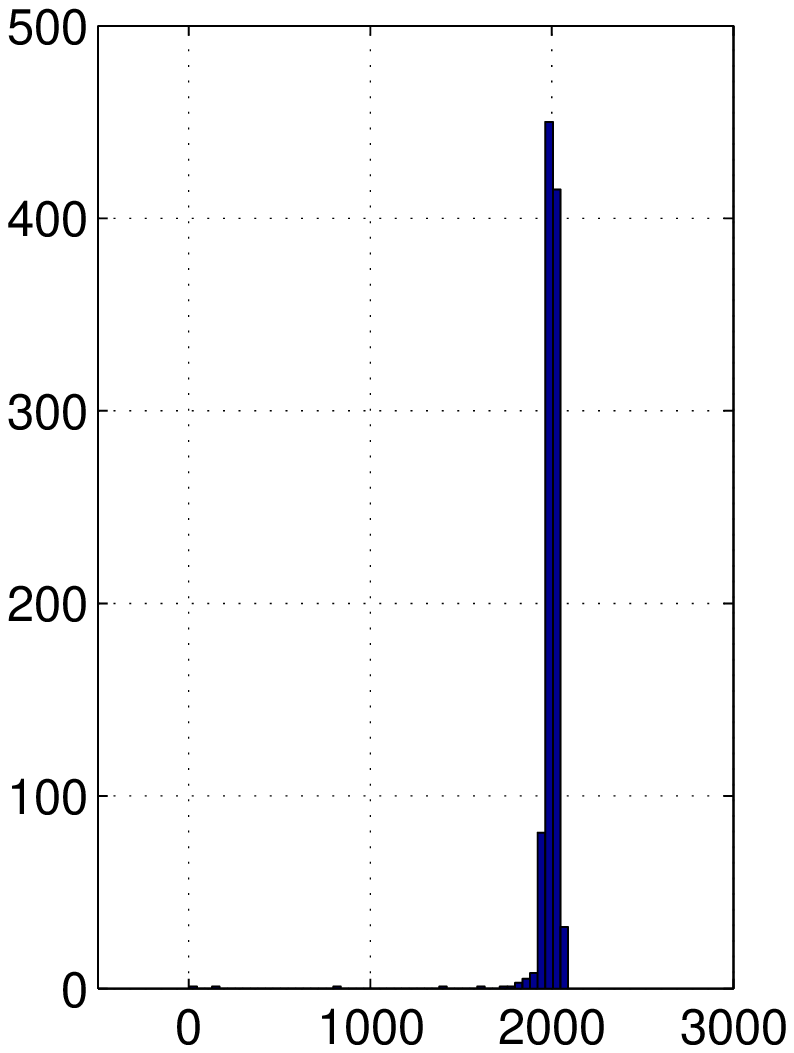} \\
\etabu\ecc
\caption{Monte-Carlo Markov Chain for the three hyperparameters generated by the proposed Gibbs sampler. From top to bottom: $\PAB$, $\PDB$ and $\PNB$. The left part of the figure shows the samples as a function of iteration index and the right part of the figure shows the samples as histograms. \label{Fig:HyperResults}}
\end{figure}

The resulting generated hyperparameters $\PAB$, $\PDB$ and $\PNB$ are shown in Fig.~\ref{Fig:HyperResults}. The left part of the figure shows the 953 iterates of the three parameters: after about 200 iterations the three parameters are stabilized and seem to be under the stationary law of the chain. The empirical mean value (approximating the \Post Mean) of the parameters respectively are $\wh{\PAB} = 2.88\,10^2$, $\wh{\PDB} = 5.91\,10^4$ and $\wh{\PNB} = 1.99\,10^3$. The iterates are also shown on the right hand side of Fig.~\ref{Fig:HyperResults} as histograms: they are clearly very concentrated around the \Post Mean (with small variance), \ie the marginal law for the hyperparameters are quasi-Dirac distributions. 

Considering the numerical value, in the sense of Eq.~(\ref{Eq:LambdaSeuil}), the equivalent regularization parameter is $\wh{\lambda}=2.17\,10^{1}$ and the equivalent threshold is $\wh{s} = 6.67\,10^{-3}$. It is noticeable that the threshold value correctly split the $\overline{\Obj}_k$ in two sets (less than $2.10^{-4}$ -- greater than $3.10^{-2}$). The point is that the method automatically adjusts hyperparameters to correctly separate the $\overline{\Obj}_k$. This is a first argument in favor of the proposed strategy in order to tune the threshold of an  $\mbox{L}_2-\mbox{L}_1$ Gibbs potential. 


%
The resulting image is shown in Fig.~\ref{Fig:Resultats} (on the third column). 
The effect of deconvolution is notable on the image in gray level as well as  on the shown profile. The three objects are correctly positioned, the orders of magnitude are respected and the zero level is correctly reconstructed: it can be seen on the entire image and in particular on the shown profile. The dynamic is also correctly restored: this aspect is notable on the shown profile around the maximum  (64-th sample). The true dynamic occupies the range 0~--~1.9  whereas the dynamic of the observed data scarcely exceeds 0~--~1.4: the proposed method restores the dynamic to 0~--~1.88 that is to say 99\% of the original variation.

%
%

A global quantitative comparison has been achieved by evaluating ($i$) the distance between original image $\Objb^\star$ and observed data $\Datab$ and ($ii$) the distance between original image $\Objb^\star$ and estimated image $\wh{\Objb}\PostMean$. The considered distances are normalized L2 and L1 distances. The main results are listed in Table~\ref{Tab:ErrMesures}, first and second columns and show an improvement by  a factor 2.95 (11.62\% to 3.93\%) for L2 distance and a factor 1.82 (35.47\% to 19.47\%) for L1 distance. 

In order to deepen the numerical study, a second estimate has been computed: $\wh{\Objb}\CondPostMean$ the Conditional \Post Mean (CPM), \ie the mean of the conditional \post  law $f_{\Objc,\Auxc|\Datac,\Cc}(\Objb,\Auxb|\Datab,\PDNAB)$. $\wh{\Objb}\CondPostMean$ is clearly a function of the hyperparameters $\PDNAB$ and a twofold evaluation is proposed. 

\bit 

\item The first estimate is the one obtained with $\PDNAB=\wh{\PDNAB}$. Practically, the marginal estimate $\wh{\Objb}\PostMean$ and the conditional estimate $\wh{\Objb}\CondPostMean(\wh{\PDNAB})$ are quasi-equal; this is due to the fact that the marginal law for the hyperparameters are quasi-Dirac distributions. Quantitatively, regarding L2 distances, the PM produces 3.93\% whereas the CPM produces 3.94\%; regarding L1 distances, the PM produces 19.47\% whereas the CPM  produces 19.50\%. In both cases, the modification is almost imperceptible.

\item The measurement of errors has also been explored for the CPM as a function $\PAB$, $\PDB$ and $\PNB$, around the \post mean $\wh{\PDNAB}$. Results are given on Fig.~\ref{Fig:MesureErreur}: in each case, smooth variation of distances is observed when varying parameters and an optimum is visible. It is reported on Table~\ref{Tab:ErrMesures} and shows almost imperceptible modification: optimization of the hyperparameters (based on the true image $\Objb^\star$) allows negligible improvement (smaller than 0.1\% for L2 error and smaller than 0.5\% for L1 error). So, the main conclusion is that, the unsupervised proposed approach is a relevant tool in order to tune parameters: it works (without the knowledge of the true image) as well as an optimized approach (based on the knowledge of the true image).

\eit


Finally, a third estimate has been computed: the Maximum \Apost (MAP). It has been computed for the LogErf and the Huber potentials. Both of them have been computed with equivalent hyperparameters (given above): $(\wh{\PAB},\wh{\PDB},\wh{\PNB})$ for the LogErf potential and $(\wh{\lambda},\wh{s})$ for the Huber potential. The two MAP solutions (LogErf and Huber) are visually indiscernible: this is expected from so similar potential. The results are presented in Fig.~\ref{Fig:Resultats}, right column: the estimated map suffers from cross-like artifact, due to the cross-like structure of the neighborhood system. Quantitatively speaking, the measurements of errors are given on Table~\ref{Tab:ErrMesures}: LogErf and Huber produce almost similar errors. Moreover, the errors are greater than the one produced by the PM and the CPM. 

The restoration is nevertheless imperfect and of limited resolution: the sharp edges remain slightly smoothed and limited in amplitude. The ringing effect also affects the quality of the deconvolved image. This diagnostic is long awaited in the framework of convex deconvolution. Anyway, the important point is not so much the property of the deconvolved image itself (intrinsic of any convex deconvolution) but the (new) practical capability to automatically tune the hyperparameters. Moreover, the potential improvement is certainly wide considering more heavy-tailed law for the auxiliary variables, as explained in the next section.

%
\section{Conclusion} \label{Sec:Conclu}

This paper presents a twofold novelty in the field of statistical image reconstruction and restoration. 
%
%
\ben

\item The partition function is explicitly given for a specific non-Gaussian Markov field, with an $\mbox{L}_2-\mbox{L}_1$ Gibbs potential. It is built as a compound field involving: an auxiliary variable following a separable Laplace distribution and a pixel variable following a Gaussian distribution given the auxiliary variable. 


\item An unsupervised deconvolution method is deduced, based on the exact likelihood taking advantage of the knowledge of the partition function. The method is fully Bayesian, and the point estimate is the posterior mean computed thanks to a Monte-Carlo Markov Chain technique. 


\een



The paper focuses on the deconvolution problem, but it is also possible to deal with simpler questions than deconvolution: parameter estimation from direct observation of the field, edge enhancement or denoising.

Moreover, the paper relies on Gaussian  noise, but the case of non-Gaussian noise is also envisaged, in particular the use of robust norms to reject abnormal data (outliers). To this end, a separable version of the $\mbox{L}_2-\mbox{L}_1$ proposed  field could be suitable as a law for noise  measurement.

The proposed method can be directly applied in the case of large support operator, \eg reconstruction problems such as Fourier synthesis~\cite{Giovannelli05}. 
The proposed methodology also remains valid for other linear model and the required modification concerns the sampling of the object. It remains Gaussian but its sampling is no longer possible in a single step for the entire image by \fftdd. The Gibbs sampling techniques constitute an adapted tool but the calculation time would be (maybe dramatically) extended. For non-linear problems, the law for the object is no longer Gaussian and a case by case study is required.

Concerning the \aprio field, other laws for auxiliary variables are certainly desirable. The possible improvements are numerous considering more heavy-tailed law in order to overcome the limitation of the convex deconvolution. The methodology still remains valid but the difficulty then concerns the sampling of the auxiliary variables. The direct sampling by inversion of the cumulative density function may not be possible, however, the rejection or the Hastings-Metropolis algorithms could be used to overcome this difficulty. 

In the case of myopic deconvolution, it is also conceivable to estimate (part of) the parameters of the observation system. Here again, a case by case study is necessary, but the delicate question of the system parameter sampling can probably be tackled by means of rejection or Hastings-Metropolis algorithms.

%
\appendices

\section{erf, erfc, erfcx} \label{Sec:AnnErf}

The $\erf$ function is defined for $x\in\eR$  by:
\beq \label{Eq:DefErf}
\Erf{x} = \frac{2}{\sqrt\pi} \int_{0}^x \Exp{-t^2} \, \dD t \,,
\eeq
and $\inverf$ denotes the reciprocal function. Elsewhere, $\Erfc{x} = 1- \Erf{x}$ and $\Erfcx{x} = \Exp{x^2} \Erfc{x}$. Concerning the latter, there are the following expansions:
\beqn
\Erfcx{x}	&\underset{+\infty}{\sim}&	\cro{1-x^{-2}} / \pth{x\sqrt{\pi}} \label{Eq:ErfcxInf}\\
\Erfcx{x}	&\underset{ 0}{\sim}&			1-\froc{2x}{\sqrt{\pi}} \label{Eq:ErfcxZero}\,.
\eeqn
and the derivative $\Erfcx{x}' = 2\,x\, \Erfcx{x} - 2/\sqrt{\pi}$.

\section{Gauss and Laplace convolution}\label{Sec:AnnL2ConvL1}

Considering the calculations, a large part of the proposed de\-ve\-lop\-ments is based on the convolution of a Gaussian function and a Laplacian function.

\subsection{Preliminary Calculi}

For  $x_0\geq 0$ and $x\in\eR$, write:
\beqx
J(x_0,x,d,b) = \int_0^{x_0} \Exp{ -  \acc{d (y-x)^2 + b y} /2 }~ \dD y \,,
\eeqx
simply written as $J(x_0,x)$ when there is no ambiguity. On rewriting the argument of the exponential, we have:
\beqx
d (y-x)^2 + b y = d \cro{ (y-\tilde{x})^2 + (x^2-\tilde{x})^2 } 
\eeqx
with  $\tilde{x} = x - b/2d$. The change of variable $t=(y-\tilde{x}) \, \sqrt{d/2}$, yields:
\beqnx
J(x_0,x)	&=& \sqrt{\froc{\pi}{2d}} ~ \Exp{ \froc{b^2}{8d} } ~ \Exp{ - b x / 2}\\
					& & \acc{ \Erf{ \tilde{x} ~ \sqrt{\froc{d}{2}} } - \Erf{ (\tilde{x}-x_0)~ \sqrt{\froc{d}{2}} } } \,,
\eeqnx
where the function $\erf$ is defined by~(\ref{Eq:DefErf}). In particular, one has: $J(0,x)=0$ and 
\beqnx
J(+\infty,x)	&=& \sqrt{\froc{\pi}{2d}} ~ \Exp{ \froc{b^2}{8d} } ~ \\
							& & \Exp{- b x / 2} ~ \Erfc{ - \tilde{x} ~ \sqrt{\froc{d}{2}} } \,.
\eeqnx

\subsection{Convolution}

For $x_0,x\in\eR$, write:
\beqx
I(x_0,x,d,b) = \int_{-\infty}^{x_0} \Exp{ -  \acc{d (y-x)^2 + b |y|} /2 } ~ \dD y \,,
\eeqx
written simply as   $I(x_0,x)$ when there is no ambiguity. By the change of variables $y'=-y$, $y'=by$, and $y'=y\sqrt{d}$, it is shown that:
\beqnx
I(+\infty,x,d,b) &=& I(+\infty,-x,d,b) \\
I(x_0,x,d,b) &=& I(b x_0,b x,d/b^2,1) \, / \,b \\
I(x_0,x,d,b) &=& I(x_0\sqrt{d},x\sqrt{d},1,b/\sqrt{d}) \,/\, \sqrt{d}
\eeqnx

It can thus be deduced that:
\beqnx
I(x_0,x,d,b) &=& J(+\infty,-x,d,b) - J(-x_0,-x,d,b)\\
I(x_0,x,d,b) &=& J(+\infty,-x,d,b) + J(x_0,x,d,b)
\eeqnx
respectively for  $x_0<0$ and $x_0\ge0$. These relationships are useful for the study of the potential function (next Appendix) and for the inversion of the cdf of $\Auxc|\Objc$ (Appendix~\ref{Sec:AnnInvCDF}).

\section{Log-Erf Potential function} \label{Sec:AnnLogErf}

According to the results of the previous Appendix the potential function of the marginal field $\Objc$, Eq.~(\ref{Eq:PotentielLogErf}), Section~\ref{Sec:CasLaplace} is written:
\beqx
\phi(x)  = -2 \log I(+\infty, x , \PDB, \PAB ) \,.
\eeqx
By putting: $\chi(x)= \Exp{\PAB \,x /2 } ~ \Erfc{ (\rho+x) ~\sqrt{\froc{\PDB}{2}} }$,  $\rho=\PAB/2\PDB$ the potential function can be written:
\beqx
\phi(x)  = -2 \Log{ \chi(x) + \chi(-x)} \,,
\eeqx
up to additive constants. The derivation shows that :
\beqnx
\chi'(x)		&=& \frac{\PAB}{2} \, \chi(x) - \sqrt{\frac{2\,\PDB}{\pi}} \, \Exp{ - \PDB \pth{\rho^2 + x^2} /2 } \\
\chi''(x)	&=& \frac{\PAB}{2} \, \chi'(x) + \sqrt{\frac{2\,\PDB^3}{\pi}} \, x \,\Exp{ - \PDB \pth{\rho^2 + x^2} /2 }
\eeqnx
and it can easily be deduced that:
\beqx
\phi'(x)  = - 2 ~ \frac{\chi'(x) - \chi'(-x)}{\chi(x) + \chi(-x)} = - \PAB ~ \frac{\chi(x) - \chi(-x)}{\chi(x) + \chi(-x)} 
\eeqx
and in particular
\beqx
\phi'(0)	= 0  ~~~\AND~~~ \phi'(+\infty)	= \PAB \,.
\eeqx
Moreover, concerning the second derivative at origin
\beqnx
\phi''(0)	&=& -\, 2 ~ \frac{\chi''(0)}{\chi(0)} = - \,\PAB ~ \frac{\chi'(0)}{\chi(0)}  \\
				&=& \frac{\PAB^2}{2} ~  \cro{ \pth{\eta \, \sqrt{\pi} ~\Erfcx{\eta}}\pmu -1 }
\eeqnx
with  $\eta = \PAB / \sqrt{8\PDB}$.

\section{Gamma Probability Density Function}\label{Sec:AnnGamma}

The  gamma probability density function  is parametrized by $\alpha>0$ and $\beta>0$ in the form:
\beq\label{Eq:GammaPDF}
f_\gamma \pth{ x \,;\, \alpha, \beta} 
	= \frac{1}{\beta^\alpha \, \Gamma[\alpha]} ~ x^{\alpha-1} ~ \Exp{-x/\beta}  \, \, \unbb_{\eR_{+}}(x) \,,
\eeq
where $\unbb_{\eR_{+}}$ is the indicator function of $\eR_+$. The expected value is $\alpha \, \beta$, the variance is $\alpha \, \beta^2$ and it is maximal for  $x=\beta(\alpha-1)$ in the case  $\alpha>1$.

\section{Integration of Hyperparameter}\label{Sec:AnnIntegreHyper}

\subsection{Preliminary Result}

Given a function $f:\eR\rightarrow\eR_+$, $\Cc^\infty$ and assume that  $g(x)=xf(x)$ can be integrated. By integrating from $0$ to $M$ the Taylor expansion of $g(x)$ at origin, one shows that: 
\beq \label{Eq:Limite}
\frac{1}{M^2} ~ \int_0^M x\,f(x)\, \dD x  \underset{\scriptscriptstyle M=0}{\longrightarrow}  \frac{1}{2} ~f(0) \,.
\eeq

Then, give a function $\psi:\eR^Q\times\eR\rightarrow\eR_+$ such that $\eps\,\psi(\ub,\eps)$ can be integrated over $\eR^{Q+1}$. By using~(\ref{Eq:Limite}), it can be seen that:
\beq \label{Eq:LimiteRapport}
\frac{\displaystyle \int_0^M \eps ~ \psi(\ub,\eps) \,\dD \eps}{\displaystyle \int_\vb  \int_0^M \eta ~ \psi(\vb,\eta) \,\dD \eta \,\dD \vb} ~
\underset{\scriptscriptstyle M=0}{\longrightarrow}  
\frac{\psi(\ub,0)}{\displaystyle \int_\vb  \psi(\vb,0) \ \dD \vb} ~
\eeq
provided that $\psi(\vb,0)$  can still be integrated over $\eR^{Q}$.

\subsection{Posterior Law}

The \apost law (Section~\ref{Sec:PostJoint}) for  $\Objc$, $\Auxc$, $\Cc$ and $\Ec$ given $\Datac$  (parametrized by the  coefficient $M_\eps$) is written, after simplification by $\delta' M_\eps$:
\beqx
\frac{\displaystyle \int_{\eps=0}^{M_\eps}  \psi(\ub,\eps) ~ \dD \eps}
{\displaystyle \int_{\Objb\Auxb\PDNAB} \int_{\eps=0}^{M_\eps} \psi(\vb,\eps) ~\dD \eps ~ \dD \PDNAB ~ \dD \Objb ~ \dD \Auxb}
\eeqx
where $\ub$ represents all the parameters $\Objb,\Auxb,\PDNAB$ and:
\beqx
\barr{l}
\psi(\ub,\eps)~ = \delta' ~ \PNB^{~\alpha_\nD-1+N/2} ~~ \PDB^{~\alpha_\dD-1+N/2} ~~ \PAB^{~\alpha_\bD-1+N} ~ \eps ~~~~~~ \\ ~\\ 
~~~~~~~~  \exp - \, \acc{ Q_\eps /2 ~+~ \PNB/{\beta_\nD} ~+~ \PDB/{\beta_\dD} ~+~ \PAB/{\beta_\bD} } \,.~\\~\\
\earr
\eeqx
To apply the relationship~(\ref{Eq:LimiteRapport}), it is sufficient to ensure that $\psi(\vb,0)$ can be integrated. %
Since the norms in $\eR^N$ are equivalent, $k\in\eR_+$ can be found such as $N_1(\Auxb)<kN_2(\Auxb)$ for all $\Auxb$. Thus the integrand can be majored by a Gaussian integrand and convergence  ensured if and only if $\rond{\matH}_{\scriptscriptstyle 00}\neq0$. 

In the limit, when $M_\eps\rightarrow 0$, we have the result~(\ref{Eq:TotalPosterior}).

\section{Inversion of $\Auxc|\Objc$ cdf} \label{Sec:AnnInvCDF}

The sampling of auxiliary variables (Section~\ref{Sec:PosteriorMean})  given the object is based on the inversion of cdf of $\Auxc|\Objc$. For $u\in[0,1]$:
\beq \label{Eq:Repart}
u = F_{\Auxc|\Objc}(\aux) = \int_{-\infty}^{\aux} f_{\Auxc|\Objc} = \frac{I(\aux,\bar{\aux},\PDB,\PAB)}{I(+\infty,\bar{\aux},\PDB,\PAB)} \,.
\eeq
is to be resolved. In order to solve this equation, write $\rho=\PAB/2\PDB$
\beqnx
\theta_- &=& \Exp{+\PAB \, \bar{\aux} /2 } ~ \Erfc{ (\rho+\bar{\aux}) ~\sqrt{\froc{\PDB}{2}} } \\
\theta_+ &=& \Exp{-\PAB \, \bar{\aux} /2 } ~ \Erfc{ (\rho-\bar{\aux}) ~\sqrt{\froc{\PDB}{2}} }
\eeqnx
and $\theta = \theta_- + \theta_+$. Moreover $s=F_{\Auxc|\Objc}(0)=\theta_- / \theta$ and $\bar{u} = \theta u - \theta_-$. Equation~(\ref{Eq:Repart}) is resolved differently depending on whether  $u\leq s$ (\ie $b\leq0$) or  $u\geq s$ (\ie $b\geq0$) and yields: 
\beqnx
b = \bar{\aux} + \rho - \InvErf{T_-} \sqrt{2/\PDB} &\IF& u\leq s\\
b = \bar{\aux}  -\rho - \InvErf{T_+} \sqrt{2/\PDB} &\IF& u\geq s
\eeqnx
where  $\InvErf{\cdot}$ is defined in Appendix~\ref{Sec:AnnErf} and
\beqnx
T_-&=& \Erf{ (\bar{\aux}+\rho) ~ \sqrt{\froc{\PDB}{2}} } - \bar{u} ~ \Exp{- \PAB \, \bar{\aux} /2 }\\
T_+&=& \Erf{ (\bar{\aux}-\rho) ~ \sqrt{\froc{\PDB}{2}} } - \bar{u} ~ \Exp{+ \PAB \, \bar{\aux} /2 } \,.
\eeqnx
Thus it is possible to sample $\Auxc|\Objc$ simply from $u$ uniformly distributed over  $[0,1]$.

\section{Conditional posterior law for $\Objc$} \label{Sec:AnnObjCondPost}

The posterior law $(\Objc,\Auxc,\Cc|\Datac$) given by Eq.~(\ref{Eq:TotalPosterior}) in Section~\ref{Sec:TotalPost} involves 	
\beqx
Q_0  = \PNB \, N_2(\Datab - \MatHb \star \Objb)	 + \PDB \, N_2(\MatDb \star \Objb - \Auxb) +  \PAB \, N_1(\Auxb) \,,
\eeqx
and the conditional posterior law $(\Objc|\Auxc,\Cc,\Datac$) required to sample object in Section~\ref{Sec:SampleObject} involves 	
\beqx
Q_{0|}  = \PNB \, N_2(\Datab - \MatHb \star \Objb)	 + \PDB \, N_2(\MatDb \star \Objb - \Auxb) \,.
\eeqx
In the Fourier domain:
\beqnx
Q_{0|}  &=& \PNB \, N_2(\rond{\Datab} - \rond{\MatHb} \Schur \rond{\Objb})	 + \PDB \, N_2(\rond{\MatDb} \Schur \rond{\Objb} - \rond{\Auxb}) \\
        &=& \sum_{pq} \PNB \,|\rond{\data}_{pq} - \rond{\matH}_{pq}  \rond{\obj}_{pq}|^2+ \PDB \, |\rond{\matD}_{pq}  \rond{\obj}_{pq} - \rond{\aux}_{pq}|^2 
\eeqnx
that is to say a separable summation. Moreover, it can be rewritten and identified to a sum of quadratic terms: 
\beqnx
Q_{0|}  &=& \sum_{pq} \rond{\nu}_{pq} |\rond{\obj}_{pq} -\rond{\mu}_{pq}|^2 
\eeqnx
with $\rond{\nu}_{pq}$ and $\rond{\mu}_{pq}$ given in Eq.~(\ref{Eq:MuNuNU})\,-\,(\ref{Eq:MuNuMU}).

\section{Empirical Least Squares Hyperparameters}\label{Sec:AnnInitHyper}

The initialization of the algorithm is based on second order statistics of the analyzed data, in the Fourier domain. Considering the structure of the \aprio field and the noise, for all $(p,q)$, such as $\rond{\matF}_{pq}\neq0$ one has:
\beqx
\rond{\Obj}_{pq} = \sqrt{\PD} ~ \rond{G}_{pq} / \rond{\matF}_{pq} 
~~~\AND~~~
\rond{\Bruit}_{pq} = \sqrt{\PN} ~ \rond{G}'_{pq}\\ 
\eeqx
where $\rond{G}_{pq}$ and $\rond{G}'_{pq}$ are two independent zero-mean white Gaussian noise with unitary variance. 
Moreover, considering the observation equation~(\ref{Eq:Convol}), one also has $\rond{\Data}_{pq} = \rond{\matH}_{pq} \rond{\Obj}_{pq} + \rond{\Bruit}_{pq}$. With $\rond{Z}_{pq}=|\rond{\Data}_{pq}|^2$ and $\rond{r}_{pq} = |\rond{\matH}_{pq}| / |\rond{\matF}_{pq}|$, we have:
\beqx
\Esp{\rond{Z}_{pq}} = \PD~ \rond{r}_{pq} + \PN \,.
\eeqx
Thus the parameters $\PD$ and $\PN$ can be selected at the minimum of the least squares criterion:
\beqx
J(\PD,\PN) = \sum_{pq} (\rond{z}_{pq} - \PD ~ \rond{r}_{pq} + \PN)^2 \,.
\eeqx
It is found that:
\beqx 
\PD = \frac{\gamma-\alpha \delta}{\beta-\alpha^2} ~~~\AND~
\PN = \frac{\beta\delta-\alpha\gamma}{\beta-\alpha^2}
\eeqx
with $N\alpha=\sum \rond{r}_{pq}$, $N\beta=\sum \rond{r}_{pq}^2$, $N\gamma=\sum \rond{r}_{pq}\rond{z}_{pq}$, and $N\delta=\sum \rond{z}_{pq}$. These values for $\PD$ and $\PN$ are used to initialize the proposed algorithm (Section~\ref{Sec:PosteriorMean}): $\PDB=1/\PD$ and $\PNB=1/\PN$. The third parameter $\PAB$ is initialized at the critical value: $\PAB = \sqrt{2\pi\PDB}$.

%
%

%
\bibliographystyle{IEEEtran}


\begin{biography}
[{\includegraphics[width=1in,height=1.25in,clip,keepaspectratio]{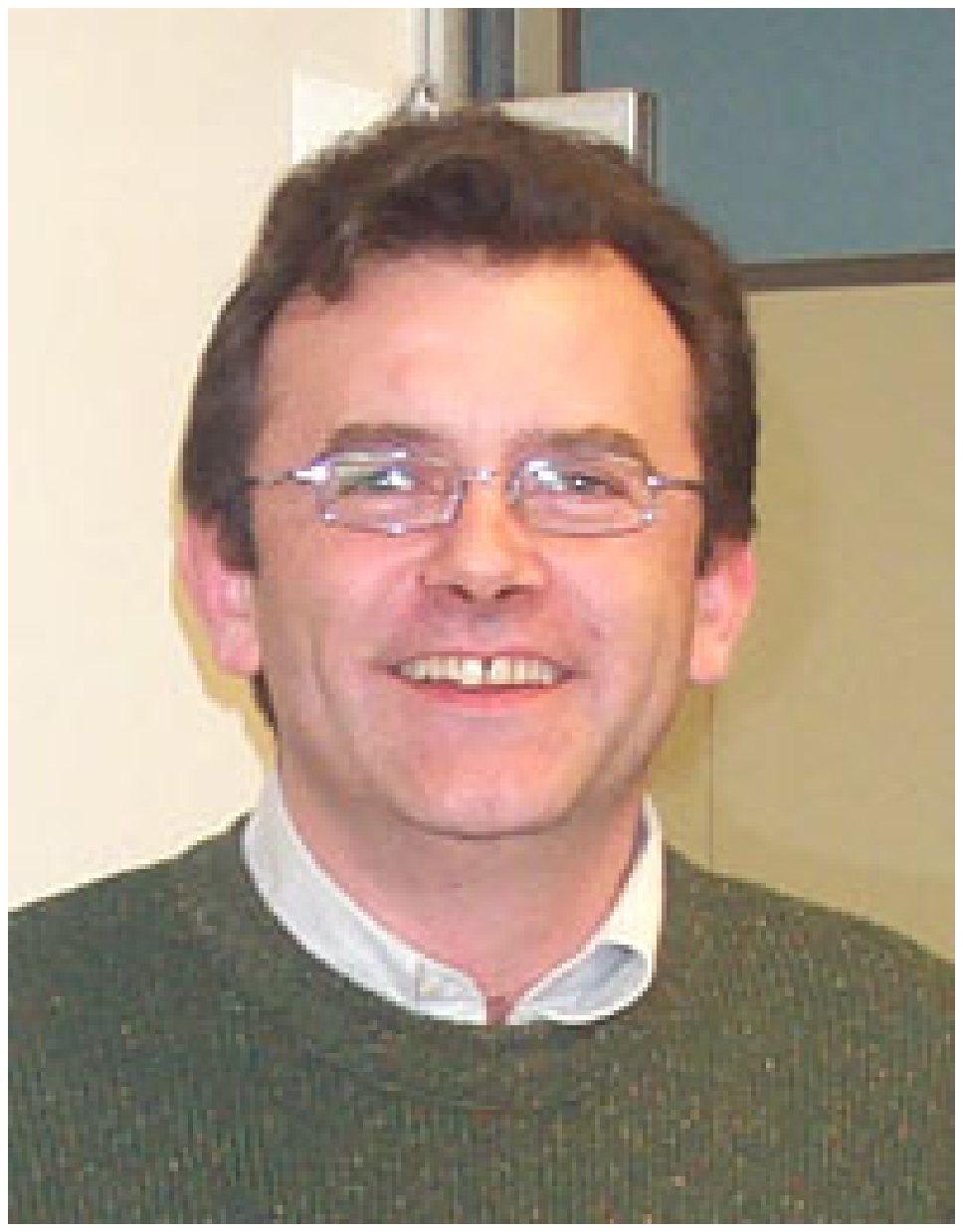}}]
{Jean-Fran\c{c}ois Giovannelli} was born in B\'eziers, France, in 1966. He graduated from the \'Ecole Nationale Sup\'erieure de l'\'Electronique et de ses Applications in 1990. He received the Doctorat degree in physics at Universit\'e Paris-Sud, Orsay, France, in 1995. \\
\indent
He is presently assistant professor in the D\'epartement de Physique at Universit\'e Paris-Sud and researcher with the Laboratoire des Signaux et Syst\`emes (CNRS - Sup\'elec - UPS). He is interested in regularization and Bayesian methods for inverse problems in signal and image processing. Application fields essentially concern astronomical, medical and geophysical imaging. 
\end{biography}

\end{document}